\documentclass[preprint,11pt]{aastex}

\shorttitle{VLA Protostars}
\shortauthors{Tobin et al.}

\newcommand{\kms}{\mbox{km s$^{-1}$}}

\begin{document}

\title{The VLA Nascent Disk And Multiplicity (VANDAM) Survey of Perseus Protostars. Resolving the Sub-Arcsecond Binary System in NGC 1333 IRAS2A}
\author{John J. Tobin\altaffilmark{1,}\altaffilmark{13,}\altaffilmark{14}, Michael M. Dunham\altaffilmark{2}, Leslie W. Looney\altaffilmark{3},
Zhi-Yun Li\altaffilmark{4}, Claire J. Chandler\altaffilmark{5}, Dominique Segura-Cox\altaffilmark{3}, Sarah I. Sadavoy\altaffilmark{6},
Carl Melis\altaffilmark{7}, Robert J. Harris\altaffilmark{3}, Laura M. Perez\altaffilmark{5,}\altaffilmark{15}, Kaitlin Kratter\altaffilmark{8}, Jes. K. J{\o}rgensen\altaffilmark{9,10}, 
Adele L. Plunkett\altaffilmark{11}, Charles L. H. Hull\altaffilmark{12}}

\altaffiltext{1}{National Radio Astronomy Observatory, Charlottesville, VA 22903}
\altaffiltext{2}{Harvard-Smithsonian Center for Astrophysics, Cambridge, MA 02138}
\altaffiltext{3}{Department of Astronomy, University of Illinois, Urbana, IL 61801}
\altaffiltext{4}{Department of Astronomy, University of Virginia, Charlottesville, VA 22903}
\altaffiltext{5}{National Radio Astronomy Observatory, Socorro, NM 87801}
\altaffiltext{6}{Max-Planck-Institut f\"ur Astronomie, D-69117 Heidelberg, Germany}
\altaffiltext{7}{Center for Astrophysics and Space Sciences, University of California, San Diego, CA 92093}
\altaffiltext{8}{University of Arizona, Steward Observatory, Tucson, AZ 85721}
\altaffiltext{9}{Niels Bohr Institute, University of Copenhagen,
  Juliane Maries Vej 30, DK-2100 Copenhagen {\O}., Denmark}
\altaffiltext{10}{Centre for Star and Planet Formation \& Natural History Museum of Denmark, University of Copenhagen, {\O}ster Voldgade 5--7, DK-1350 Copenhagen
  {K}., Denmark}
\altaffiltext{11}{Department of Astronomy, Yale University, New Haven CT 06520}
\altaffiltext{12}{Astronomy Department \& Radio Astronomy Laboratory, University of California, Berkeley, CA 94720}
\altaffiltext{13}{Current Address: Leiden Observatory, Leiden University, PO Box 9513, 2300 RA, Leiden, The Netherlands; tobin@strw.leidenuniv.nl}
\altaffiltext{14}{Hubble Fellow}
\altaffiltext{15}{Jansky Fellow of the National Radio Astronomy Observatory}

\begin{abstract}
We are conducting a 
Jansky VLA Ka-band (8 mm and 1 cm) and C-band (4 cm and 6.4 cm) 
survey of all known protostars in the Perseus Molecular Cloud, providing resolution down to
$\sim$0\farcs06 and $\sim$0\farcs35 in Ka-band and C-band, respectively. 
Here we 
present first results from this survey that enable us
to examine the source NGC 1333 
IRAS2A in unprecedented detail and resolve it into a proto-binary system separated by
 0\farcs621$\pm$0\farcs006 ($\sim$143 AU) 
at 8 mm, 1 cm, and 4 cm. These 2 sources (IRAS2A VLA1 and VLA2) are likely driving the two 
orthogonal outflows known to originate from IRAS2A. 
The brighter source IRAS2A VLA1 is extended perpendicular to its outflow
in the VLA data, with a deconvolved size of 0\farcs055 ($\sim$13 AU), 
possibly tracing a protostellar disk. The recently reported candidate companions
(IRAS2A MM2 and MM3) are not detected in either our VLA data, CARMA 1.3 mm data, or SMA 850 \micron\
data. SMA CO ($J=3\rightarrow2$), CARMA CO ($J=2\rightarrow1$), 
and lower resolution CARMA CO ($J=1\rightarrow0$) observations
are used to examine the outflow origins and the nature of the candidate companions to IRAS2A VLA1. 
The CO ($J=3\rightarrow2$) and ($J=2\rightarrow1$) data show that IRAS2A MM2
is coincident with a bright CO emission spot in the east-west outflow, and IRAS2A MM3 is within the north-south outflow.
In contrast, IRAS2A VLA2 lies at the east-west outflow symmetry point. We 
propose
that IRAS2A VLA2 is the driving source of the East-West outflow and a true companion
to IRAS2A VLA1, whereas IRAS2A MM2 and MM3 may not be protostellar.

\end{abstract}

\keywords{ISM: individual (NGC 1333) --- planetary systems: proto-planetary disks  --- stars: formation}

\section{Introduction}

The formation of binary and multiple star systems is a common outcome of the star formation
process. The multiplicity frequency of field stars is a strong function of spectral type, 
with higher fractions of multiple systems in early-type stars compared to late-type stars.
For example, nearly all O stars, about half of all solar-type stars, and about 
35\% of M stars are multiples, with multiplicity declining into the brown dwarf regime
\citep{sana2011, raghavan2010, lada2006, duchene2013}. 
The even higher multiplicity fractions observed in young star forming regions like Taurus 
suggest that binary formation is commensurate with star formation \citep{kraus2011}.

Many studies have sought to characterize multiplicity during
the protostellar phase of the star formation process.
The youngest identifiable protostars are the Class 0 sources, characterized
by a dense envelope of gas and dust surrounding 
the central object
\citep{andre1993}.
Class I protostellar systems 
are more-evolved, having envelopes that are
 substantially
less optically thick and have a smaller reservoir of mass. 
The formation of a circumstellar disk also takes place in the Class 0 and I phases due to conservation of
angular momentum during collapse \citep{cassen1981}.
Finally, there are also
Class II sources, pre-main sequence stars with an infrared excess above the stellar photosphere indicative of
a proto-planetary disk, and Class III without an infrared excess. \citep{lada1987}.

\citet{looney2000} conducted one of the first large censuses of multiplicity 
toward Class 0 protostars, finding a high incidence of multiplicity toward Class 0 protostars 
on $\sim$1000 AU scales. Similarly, \citet{connelley2008} found that the more-evolved 
Class I protostars also have a high degree of multiplicity, in excess of the field stars. 
A more recent high-resolution study of 5 Class 0 
protostars by \citet{maury2010} did not detect any binary sources between
150 AU and 550 AU; these results combined with those of \citet{looney2000} and \citet{connelley2008} lead
the authors to suggest that multiplicity increased from the Class 0 to Class I phase.
On the other hand, a larger Class 0 sample examined by \citet{chen2013} found the opposite trend,
that multiplicity is highest in the Class 0 phase and decreases with 
evolution \citep[see PPVI review][]{reipurth2014}. Despite 
recent progress, the \citet{chen2013} sample was limited by non-uniform 
resolution, non-uniform sensitivity, and a large mean linear resolution of 600 AU, which  
is much larger than the median field star separation of $\sim$50 AU. These limitations make 
inferences on the origin of multiplicity difficult from the current data.

The dominant modes of multiple system formation are (1) the fragmentation of 
the core or envelope during collapse due to turbulence and/or rotation in the envelope 
\citep[e.g.,][]{inutsuka1992,bb1993,offner2010,boss2013,boss2014} and (2) the fragmentation of a rotationally 
supported accretion disk via gravitational instability 
\citep[e.g.,][]{bonnell1994a,bonnell1994b,kratter2010,vorobyov2010a}.
Disk fragmentation will preferentially form relatively close binaries, while 
turbulent/rotational fragmentation can form relatively wide multiples that may (or may not)
evolve to smaller separations.
Both routes point to formation early in the star formation process 
when there is still a significant mass reservoir available.
Since Class 0 sources have large massive envelopes that obscure most radiation
shortward of 10 \micron, high resolution millimeter/centimeter 
imaging is needed to reveal multiple systems.

To characterize the multiplicity of protostellar systems in a systematic manner, we are using 
the Karl G. Jansky Very Large Array (VLA) to conduct the VLA Nascent Disk and Multiplicity 
survey (VANDAM) toward all known protostars in the Perseus molecular cloud. Perseus is one of the
nearest star forming regions at a distance of $\sim$230 pc \citep{hirota2008,hirota2011}. This survey is being conducted
at four wavelengths: 8 mm/1 cm (Ka-band) and 4 cm/6.6 cm (C-band). The best resolution of the survey is 
$\sim$0\farcs06 (14 AU) in Ka-band, with observations in both A and B configurations. 
The C-band data were taken in only A configuration, yielding $\sim$0\farcs35 (80 AU) resolution.
The Ka-band observations toward protostars are typically sensitive to a combination of thermal dust 
emission and thermal free-free emission from protostellar jets, and the C-band data are 
sensitive to thermal free-free emission and non-thermal
synchrotron emission (if present) \citep{anglada1995}.
We observed all the reported protostars in the \citet{enoch2009} \textit{Spitzer Space Telescope} survey in addition to
millimeter-detected sources \citep{enoch2010,chen2010, pineda2011,schnee2012,hirano1999} and some
far infrared sources detected by the \textit{Herschel Space Observatory} Gould Belt Survey 
of Perseus \citep[e.g.,][]{pezzuto2012,sadavoy2014}.

Many of the best studied and most luminous protostars are located in the NGC 1333 sub-region of Perseus. 
NGC 1333 IRAS2A (hereafter IRAS2A), in particular, has received intensive study of its multiplicity \citep{looney2000,maury2010},
outflows \citep{sandell1994,engarg1999,sandellknee2001,wakelam2005,plunkett2013,codella2014,walkersmith2014},
chemistry \citep{jorgensen2004b,jorgensen2005,maury2014}, and kinematics \citep{brinch2009,maret2014}.
\citet{sandell1994} found two nearly orthogonal bi-polar outflows apparently originating from IRAS2A in a
single-dish CO map, and \citet{engarg1999} more clearly resolved the outflows
with a BIMA CO ($J=1\rightarrow0$) map of the region. The two outflows have long been regarded
as evidence for a close multiple \citep{jorgensen2004a} given the similarities
to other multiple outflow proto-binaries \citep[e.g., L723, ][]{carrasco2008,girart2009}. 
While several high-resolution studies failed to identify a close companion 
\citep{looney2000,maury2010}, recent Plateau de Bure Interferometer (PdBI) 
observations of IRAS2A detected 
the main protostellar source and
two candidate companion
sources in maps of 1.4 mm dust continuum emission, denoted MM1, MM2, and MM3 \citep{maury2014,codella2014}.
MM1 is the brightest source and the candidate companions to MM1 
are separated by 2\farcs4 (MM2) and 2\farcs5 (MM3), respectively. 
Although both MM2 and MM3 were detected at 1.4 mm, only MM2 was also detected at 3.2 mm.  
Thus, \citet{codella2014} considered MM2 as protostellar and MM3 as a possible outflow feature.

In this paper, we present new VLA Ka-band and C-band continuum 
observations toward IRAS2A, with unprecedented angular
resolution and sensitivity. We detect a new component separated
from the main source by $<$ 1\arcsec\ in our VLA data at 8 mm, 1 cm, and 4 cm. In contrast,
 neither MM2 nor MM3 are detected in our VLA data or complementary 1.3 mm and 850 \micron\ imaging. We describe the
observations in Section 2, the continuum imaging results are discussed in Section 3, the flux densities
and spectral indices in Section 4, and discuss the sources in the context of the
outflows in Section 5. The results are discussed in Section 6, and our 
conclusions are summarized in Section 7.

\section{Observations}
\subsection{VLA Observations}
We observed IRAS2A with the VLA in Ka-band in B and A configuration on 
2013 November 04 and 2014 February 21, respectively. The Ka-band observations 
were taken in 3-bit mode, providing 8 GHz of bandwidth
that are divided into 64 sub-bands, each with 128 MHz bandwidth and 2 MHz channels
and full polarization products. We centered one 4 GHz baseband at 36.9 GHz and the other
at 28.5 GHz. IRAS2A was observed 
in B configuration within two 3.5 hour scheduling blocks (one pre-transit and one post-transit), 
sharing a track with two other sources. The A configuration data were obtained in a 3 hour
scheduling block shared with only one other source.
3C48 and 3C84 were observed as the absolute flux and bandpass calibrators, 
respectively. 
The science observations were conducted in fast-switching mode to compensate for the rapid atmospheric 
phase variations, alternating between the science target and gain calibrator (J0336+3218) 
with a total cycle time of 2.5 minutes; pointing solutions were
updated once every 50 minutes. The total time on source was
 $\sim$60 minutes in B-configuration and $\sim$35 minutes in A-configuration.

The C-band data were taken on 16 March 2014 in 8-bit mode, 
yielding 2 GHz of bandwidth divided into
sixteen 128 MHz sub-bands with 2 MHz channels and full polarization.
We centered one baseband at 4.7 GHz and the other at 7.3 GHz.  
This setup avoids RFI between 4.0 GHz and 4.2 GHz and sensitivity degradation 
near the band edges. The scheduling
block was 3 hours long and shared observations with another field. We observed 3C48 as both 
the absolute flux density and bandpass calibrator for 10 minutes, then the gain calibrator, J0336+3218,
for 1 minute and the science fields for 9 minutes. The total time on source for each science field
was $\sim$60 minutes. Pointing updates were not necessary at C-band due to the large primary beam.

The data were reduced using CASA 4.1.0 and version 1.2.2 of the VLA pipeline.
Flags applied include sub-band edges and the online flags generated by the VLA system. We carried out additional 
data editing by examining the gain, phase, and bandpass calibration solutions. We flagged the calibration
solutions that did not follow expected trends and then applied these flags to the calibrated measurement set using
the \textit{applycal} task in \textit{flagonly} mode, which flags the science data without accompanying or flagged
calibration solutions. The data were imaged using the \textit{clean} task in multi-frequency synthesis mode
with a mask drawn close to the source emission in the dirty map down 
to a level twice the rms noise. 
When imaging the full bandwidth of the two widely separated basebands in Ka-band and C-band
 we used the nterms = 2 option for \textit{clean}. With this option, the sky brightness
is modeled as a linear combination of Gaussian-like functions with amplitudes defined by Taylor-polynomials 
as a function of frequency. For imaging of the individual basebands  (4 GHz for Ka-band and
1 GHz for C-band) we only used nterms = 1 due to the lower fractional bandwidth.
Imaging was performed iteratively, cleaning and rechecking
the image for new sources apparent after cleaning the brightest sources; this step was especially
important for the C-band data which had numerous extragalactic sources dominating the dirty maps.
The absolute flux calibration uncertainty of the VLA datasets is $\sim$10\%.

\subsection{CARMA Observations}

We observed IRAS2A with the Combined Array for Research in Millimeter-wave Astronomy (CARMA), located
in the Inyo Mountains of California in B and C array configurations. The C-array data
were taken on 2013 October 20 and 21, and the B-array data were taken on 2013 December 17.
The B-array data were taken 
during a move to a more compact configuration with only with 8 antennas 
operating (6 - 10.4 m and 2 - 6.1 m). The central frequency was 225 GHz, the 
correlator was configured for single polarization, and all 8 windows were
configured for 500 MHz continuum with 15 channels each. The flux calibrator was MWC349, and 3C84 was
both the bandpass and gain calibrator. The C-array data were taken in dual-polarization mode
with a single 500 MHz continuum window and the other three windows configured 
for spectral line observations; in dual-polarization mode the
total number of independent spectral windows is halved. The spectral lines observed were 
$^{12}$CO, $^{13}$CO, and C$^{18}$O ($J=2\rightarrow1$), but only the $^{12}$CO data are discussed
in this paper. The C-array data observed Uranus as the flux calibrator and
3C84 as both the bandpass and gain calibrator. The data were edited, calibrated, 
and imaged using the Multichannel Image Reconstruction, Image Analysis, and Display (MIRIAD) software package \citep{sault1995}; see \citet{tobin2013} for further details on CARMA data reduction.
The absolute flux calibration uncertainty of the CARMA datasets is $\sim$10\% - 20\%.

We also utilized subsets of two archival CARMA D and E-array datasets in this study: 1.3 mm continuum 
and CO ($J=2\rightarrow1$) data obtained as part of the TADPOL project and 
described in \citet{hull2014} and 3 mm CO ($J=1\rightarrow0$) 
data presented by \citet{plunkett2013}. The observations and data reduction are discussed
in the relevant publications.

\subsection{SMA Observations}

Observations of IRAS2A in CO ($J=3\rightarrow2$) line emission and 850 \micron\ 
continuum were obtained with the 
Submillimeter Array \citep[SMA,][]{ho2004} in both the compact and extended 
configurations on 2004 October 17 and 2006 January 08, respectively, providing 
projected baselines ranging from approximately $5-220$ m. These data were
previously published in \citet{jorgensen2005} and \citet{brinch2009} for the compact and extended
configurations, respectively. The correlator 
was configured to provide 256 channels in the 104 MHz band, providing a channel 
separation and total bandwidth of 0.35 km s$^{-1}$ and 90 km s$^{-1}$, respectively.  
Bright, nearby quasars were used for bandpass and gain calibration following 
standard calibration procedures, and the absolute flux calibration is accurate 
to $\sim$20\%.  The data were inspected, flagged, and calibrated using the MIR 
software package\footnote{Available at https://www.cfa.harvard.edu/$\sim$cqi/mircook.html} 
and imaged, cleaned, and restored using the MIRIAD software package configured for the 
SMA\footnote{Available at http://www.cfa.harvard.edu/sma/miriad/}.

\section{Continuum Imaging Results}

The continuum emission from IRAS2A is detected at both Ka-band and C-band.
Figure 1 shows the detection emission toward IRAS2A for Ka-band (9 mm) and 
C-band (4 cm and 6.4 cm) from the A-configuration observations or the combined 
A and B configuration data. Additionally, Figure 1 also includes 
a zoom-in of the 9 mm image using only A-configuration data.
Both 9 mm images show a clear detection of a secondary source 
separated by 0\farcs621$\pm$0\farcs006 ($\sim$143 AU), and this secondary source
is detected with a signal-to-noise ratio (SNR) of 41 (A+B combined image); see Table 1. We
designate the brighter source as IRAS2A VLA1 (hereafter VLA1) and the new, fainter source 
as IRAS2A VLA2 (hereafter VLA2). The positions of VLA1 and VLA2 are 
03:28:55.5695, +31:14:37.020 and
03:28:55.563, +31:14:36.41, respectively, and are derived from 
Gaussian fits to the A+B combined 9 mm image shown in Figure 1. The 4 cm image also
shows the two sources, though the secondary source is only detected with SNR $\sim$5 and it is located within 1/4 beam of
the Ka-band position. VLA1 is detected at 6.4 cm, but VLA2 is undetected at 6.4 cm. The non-detection
of VLA2 at 6.4 cm is consistent with a declining spectral index 
(see Section 4).

The 9 mm images show extended structure in the vicinity of VLA1. The
A+B combined image shows some faint diffuse emission extending northwest of VLA1,
and the A-array image shows that the source appears extended perpendicular to 
its outflow direction (see Section 5 for discussion of the outflows). 
Gaussian fits to the 9 mm image of VLA1 yields a deconvolved position angle of 111.6\degr,
nearly perpendicular to the outflow (see Table 1); the 8 mm and 1 cm 
images alone also have consistent position
angles. Thus, in addition to resolving a close
companion source (VLA2), we are also resolving compact structure toward VLA1. 
VLA2 does not appear to have significant resolved structure with our current resolution.

The main continuum source, VLA1, is spatially coincident
with the brightest millimeter source MM1 as identified by \citet{codella2014}.
The secondary sources detected by these authors (MM2 and MM3), however, are not
detected in either Ka-band or C-band.  
\citet{codella2014} measured MM2 to have a peak 1.4 mm intensity of 15 mJy/beam 
and to be unresolved; they also claim a detection at 3.2 mm of 2 mJy/beam giving a spectral 
index ($\alpha$) of 2 to 2.5 
(we will use the convention of $S_{\lambda}$ $\propto$ $\lambda^{-\alpha}$
throughout this paper). 
However, their 3.2 mm map suggests that this 3.2 mm 
detection is tentative.  Extrapolating the MM2 flux densities to 8 mm with $\alpha$ = 2.0 (2.5), 
the expected peak intensity is 0.46 (0.19) mJy/beam (assuming unresolved 
emission). The rms noise of our Ka-band image is 8.4 $\mu$Jy/beam,
thus, MM2 should be detected with SNR = 55 (22.5) for $\alpha$ = 2.0 (2.5).  
MM3 has a peak 1.4 mm intensity of 21 mJy/beam and was not detected at 3.2 mm, 
indicating a $\alpha$ $\ga$ 4.  This source would fall below 
our sensitivity, but \citet{codella2014} did not claim a protostellar 
origin for this source, despite it being brighter than MM2.

Our non-detections with the VLA are not necessarily surprising, given that
\citet{maury2010} did not detect MM2 or MM3 at 1.3 mm in observations of 
IRAS2A with higher spatial resolution and comparable sensitivity ($\sigma$=1.16 mJy/beam).  
Furthermore, \citet{chen2013} did not detect MM2 or MM3 in 850 \micron\ 
SMA data despite having sufficient resolution and sensitivity.  The lack of 
detection in our VLA data combined with these facts prompted us to further 
investigate the nature of MM2 and MM3.  

The left panel of Figure 2 shows a 
1.3 mm CARMA continuum image of IRAS2A consisting of data from B, C, and D-arrays
with tapering and robust weighting applied.
 No point-like emission is observed toward MM2 or MM3 despite 
having sufficient sensitivity to detect them at a level better than 8$\sigma$, 
although the 1.3 mm emission is extended toward their locations.
We show a higher resolution 1.3 mm image in Figure 2 that is still 
sensitive enough to detect MM2 and MM3 at better than 10$\sigma$\footnotetext{The
lower-resolution has less sensitivity because it does not completely recover
the extended envelope emission, leading to imaging artifacts with larger amplitudes than
at higher resolution}.  These
data also have sufficient resolution to resolve VLA2 and there is emission at
the position of VLA2 at the 5$\sigma$ level, slightly extended from VLA1. 
We caution that this is a very tentative detection of VLA2
at 1.3 mm, and higher sensitivity imaging is necessary to clearly 
resolve and detect VLA2 at millimeter wavelengths.

SMA 850 \micron\ images of IRAS2A are shown in Figure 3; these images are derived from the same data 
presented by \citet[][Figure 5]{chen2013}. Like the CARMA 1.3 mm data, no
point-like emission toward the locations of MM2 and MM3 is detected. MM2 and MM3 have 
expected peak flux densities at 850 \micron\ of 40.3 mJy beam$^{-1}$ and 72 mJy beam$^{-1}$ 
respectively, assuming $\alpha$ = 2.0. In the highest-resolution image, we do 
detect extended emission at the location of MM2 with a peak intensity of $\sim$ 15 mJy/beam, but 
this is much fainter than expected from the observations of \citet{codella2014}. We also
note that \citet{chen2013} suggested a different companion source, and we do not
find evidence for this source in our VLA or CARMA data.

\section{Flux Densities and Spectral Indices}

The flux densities of IRAS2A VLA1 and VLA2 were measured using
Gaussian fits to the images at each wavelength from 850 \micron\ to
6.4 cm using the CASA task \textit{imfit} (see Table 1). 
The Gaussian fits were performed simultaneously
for the two sources at wavelengths between 1.3 mm and 4 cm; single source
fits were conducted for 850 \micron\ and 6.4 cm. For the 1.3 mm fit to VLA2, 
it was also necessary to fix the source position and
assume that it is unresolved. We also
included the 2.7 mm continuum flux from \citet{looney2000} to better constrain
the thermal dust slopes; VLA1 was detected in these data and an upper limit
is provided for VLA2. We have used these data to plot the radio spectra
of these sources in Figure \ref{sedplots}, where the Ka-band flux densities are
 taken from the A+B combined imaging. The broad wavelength coverage
is dominated by thermal dust emission at the short wavelength end and
free-free continuum at the long-wavelength end. We have attempted to decompose 
the radio spectra into their free-free and thermal dust components, assuming that
the emission spectrum of both can be described by power-laws.

We used the \textit{mpfit} routines \citep{markwardt2009} to perform 
a two-component linear fit to the data. In addition to the statistical uncertainty, 
we adopt an additional 25\% uncertainty at 1.3 mm and 850 \micron, 10\% uncertainty 
at 8 mm, 9 mm, and 1.03 cm, and 5\% uncertainty at 4 cm, 4.9 cm, and 6.4 cm. Larger uncertainties are
assumed at the shorter wavelengths because of greater calibration uncertainty and sensitivity
to dust emission from larger scales than probed by the VLA. The 850 \micron\
data will also include flux from both sources, but VLA2 is apparently much fainter than VLA1
at shorter wavelengths.

 For VLA1, where we had the most 
available data, we first simultaneously fit the data using
a two-component linear fit (method 1; described above). We also fitted the
slopes in a two step process where we used the 
4 cm and 6.4 cm data alone to derive a spectral slope
of the free-free emission (method 2; implicitly assumes no dust is contributing at those
wavelengths). We then used the free-free slope as a fixed parameter
and fit the slope of the thermal dust emission. The result from the
two-step process is shown in Figure \ref{sedplots}. We note that both methods
yield consistent slopes within the uncertainties (see Table 2). The simultaneous fits
yield shallower slopes for both the thermal dust and free-free emission than 
the two-step fits. We adopt
the free-free spectral index of 1.1 from the 4 cm to 6.4 cm fit 
(method 2) as 
the most reliable. Thus, the relative contributions of thermal dust emission to the continuum
emission at 8 mm, 9 mm, and 1.03 cm are 53\%, 43\%, and 29\% respectively.

The spectral indices of VLA2 were more difficult to derive accurately 
for both the dust and free-free components
due to weaker emission and non-detections at both the shortest and
longest wavelengths. We were only able
to perform a simultaneous fit of the thermal and free-free components because the source was not 
detected at 6.4 cm. The thermal slope is poorly constrained, and the source
appears to be dominated by free-free emission at $\lambda$ $\ga$ 2 mm. 
To better constrain the free-free slope, we assumed a thermal component with a fixed
spectral index of -3.0 (this is similar to that of VLA1 and implies a dust opacity
spectral index of 1 in the optically thin limit), comparable to previous 
observations of young stellar objects \citep{rodmann2006, melis2011, tobin2013}. 
We also fit a single slope to all the VLA data ($\lambda$ $>$ 7 mm), and we find
a slope equivalent to the free-free slope derived from the 
fit with a fixed thermal dust power-law. In both cases, the free-free emission is $>$ 50\%
of the total continuum flux to $\lambda$ $\sim$ 1 mm.

The cm-wave spectral indices ($\alpha$) are 1.1 $\pm$ 0.19 and 1.7 $\pm$ 0.15 for VLA1 and VLA2,
 respectively. These spectral indices are consistent 
with previous observations of free-free emission toward protostellar objects, with moderately optically
thick emission \citep{anglada1998,shirley2007}. The spectral indices for VLA1 and VLA2 indicate
optical depths of $\sim$0.9 and $\sim$2 at 4 cm, respectively \citep[][Equation 6]{anglada1998}.
Note that our flux densities at 4 cm for VLA1 differ by a factor of $\sim$2
from the values at 3.6 cm (0.22 mJy) given in \citet{rodriguez1999} and \citet{reipurth2002}; however,
the 6 cm flux density in \citet{rodriguez1999} is consistent with our 6.4 cm measurement. 
Spatial filtering is not likely the cause of the discrepancy at 4 cm 
because \citet{reipurth2002} also observed VLA1 in A-configuration and VLA2 is too faint 
to cause the difference. Thus, the source may be exhibiting
intrinsic variability. Due to the lower flux density at 4 cm, the spectral index we measure at centimeter wavelengths
is significantly smaller than the value of 2.6$\pm$0.4 found by \citet{rodriguez1999}.

\section{Outflow Driving Sources}

\citet{sandell1994} first observed the
 two orthogonal outflows driven from IRAS2A, but their
 single-dish maps could not identify the driving sources.
Interferometric maps of CO ($J=1\rightarrow0$) emission were observed by \citet{engarg1999},
suggesting a very close proto-binary driving the two outflows from IRAS2A.
Figure 5 shows the large-scale
CO ($J=1\rightarrow0$) outflows using the data from \citet{plunkett2013}.
These data clearly show the well-collimated east-west outflow and a wider, less-collimated
 north-south outflow. The 4.5 \micron\ emission from \citet{jorgensen2006} (grayscale in Figure 5)
agrees well with the outflow axes defined from the CO emission for the north-south flow. The
east-west flow has some tenuous 4.5 \micron\ emission associated with the blue-shifted lobe.
We adopt a position angle (PA) of 281\degr\ for the east-west outflow and a PA of 201\degr\ for the north-south
outflow, using the convention of the PA being the angle of the blue-shifted lobe east of north. 
The PA of each outflow
is defined by drawing vectors that align as closely as possible to the CO ($J=2\rightarrow1$) and CO ($J=3\rightarrow2$) 
emission maps shown in Figure 6.

The smaller-scale outflow emission is highlighted in Figure 6 using
higher resolution
CO ($J=2\rightarrow1$) and CO ($J=3\rightarrow2$) emission.
The higher resolution and higher excitation lines show that the north-south outflow 
becomes well-collimated on smaller scales, and the east-west outflow remains 
more narrow and well-collimated. VLA1 is clearly located at the base of both the blue-shifted and
red-shifted emission from the north-south outflow. Similarly, VLA 2 is located
near the origin of the east-west outflow and is 
in close proximity to
the base of the redshifted emission. However, the base of the
blue-shifted component is unclear due to blending with the blue-shifted component
of the north-south outflow. 
Imaging only the higher resolution CO ($J=3\rightarrow2$) did not conclusively
reveal a blue-shifted lobe near VLA2.
In addition to these morphological arguments, it is known that the bolometric luminosity
of a protostar is related to the outflow energy \citep{bontemps1996,wu2004}, and
the centimeter-wave luminosity (3.6 cm and 6 cm) is also correlated with L$_{\rm bol}$ and the
outflow energy \citep{curiel1989,anglada1995,shirley2007}. Furthermore, \citet{plunkett2013} found
 that the north-south outflow had 6 times more energy
than the east-west outflow. These relationships are consistent with VLA1 (brighter at all wavelengths) being
the driving source of the north-south outflow and VLA2 driving the less energetic east-west outflow.

The source MM2 from \citet{codella2014} is spatially coincident with the
peak of the red-shifted CO emission in both maps (see Figure 6). This position
is also consistent with the redshifted SiO peak and near the SO peak found by \citet{codella2014}.
The position of MM2 is inconsistent with the origin of the 
red-shifted east-west outflow, being offset to the east and slightly south of the apparent
origin in the CO emission. The non-detection of MM2 in 
any of our continuum data and placement within the east-west outflow suggests
that this source is not likely the driving source of the east-west outflow
and that it may not be protostellar in nature. Similarly, MM3 also resides
within the blue-shifted north-south outflow from IRAS2A and
is located near the peak of CO ($J=3\rightarrow2$) emission, and we agree
with \citet{codella2014} that it is not likely protostellar. 

The placement of VLA1 at the apparent north-south outflow origin coupled with its
greater centimeter luminosity indicates that it is most likely the driving source of the north-south outflow.
VLA2 is at the base of the red-shifted lobe of the
east-west outflow. The lower outflow energy of the east-west outflow and the weaker centimeter luminosity
of VLA2 makes this source the most likely candidate to drive the east-west outflow.

\section{Discussion}

The multiplicity of IRAS2A is important on its own and in the context of the origin of
protostellar multiplicity. It is often observed as a prototypical source 
for envelope structure, infall, tracers, chemistry, and outflow shocks and the multiplicity
of this source may have implications for the interpretation of these observations. %Moreover,
As we discuss below, the novel high-resolution data presented herein provide evidence for the origin of the IRAS2A system.

The VLA data have enabled us to convincingly resolve
IRAS2A into two sources separated by only $\sim$143 AU. 
At 8 mm and 1 cm, the dust is expected to be
optically thin, and the data are sensitive to a combination of dust and free-free emission,
possibly helping sources to stand out even if their dust emission is faint. 
At 4 cm the continuum should be dominated by free-free emission originating from the
base of the protostellar jets \citep{anglada1995}. IRAS2A VLA2 highlights the
utility of probing multiple emission mechanisms, because the CARMA
1.3 mm data with high enough resolution to resolve the binary show only very faint
emission toward the position of VLA2. This is an indication that this source
has much less circumstellar dust than VLA1. On the other hand VLA2 is quite bright at 8 mm and 1 cm
due to free-free emission but still fainter than VLA1. 
The detection of free-free emission is strong evidence
for the presence of a protostellar object because 
free-free emission is thought to be associated with shocks at the base of the jet, within
$\sim$10 AU of the protostar \citep{anglada1998}. However, a lack of detected free-free
emission is not evidence against a source being protostellar.

Recent PdBI results from \citet{codella2014} and \citet{maury2014} detected
two source candidates (MM2 and MM3) within 2\farcs5 of IRAS2A in dust continuum
emission at 1.4 mm. However, we did not detect these sources in our VLA data,
despite the estimated flux density of MM2 being within our sensitivity limit at 8 mm. 
Moreover, any source driving an outflow
strong enough to produce shocked emission at its position would naively be 
expected to have free-free emission associated with it at 4 cm. MM2 and MM3
were also undetected at 1.3 mm and 850 \micron. There is 3$\sigma$ emission at
the position of MM2 at 850 \micron\ that appears as part of the structure extending
from IRAS2A, and it is below the expected intensity level. Thus, it is clear that caution
should be exercised when interpreting continuum maps in which only $\sim$10\% - 20\% of the
total continuum flux is recovered, and the non-detections of MM2 and MM3 at any 
wavelength in this study indicates that these sources may be spurious. 

Numerous other studies with the PdBI also failed to detect MM2 and MM3,
despite having sufficient sensitivity and angular resolution 
\citep[][and Private Communication with those authors]{coutens2014,persson2012}. Furthermore, \citet{maury2010}
observed IRAS2A with higher resolution and sensitivity than \citet{codella2014} 
but did not detect either source. \citet{maury2010}
also reported a source $\sim$6\arcsec\ southeast of IRAS2A (called
IRAS2A SE), and we do not detect it in any of our data as well.

The fact that we see emission extended toward the positions of MM2 and MM3 at 850 \micron\
and 1.3 mm suggests that their detections could result from spatial
filtering artifacts of extended envelope
emission. For example, the uv-plane of the PdBI is sparsely sampled with at most 15 baselines
per configuration. Another possibility is that these are 
outflow-induced continuum features resulting from shock-heating
\citep[e.g.,][]{maury2010}, given that they are located near 
bright CO emission (Figure 4) and at the location of SiO and 
SO emission \citep{codella2014}. However, protostellar outflow shocks are
not expected to efficiently heat dust \citep{draine1983,hollenbach1989}. 
Nevertheless, if the features did result from outflow shock-heating we would still
expect to detect the outflow features at 1.3 mm, 850 \micron,
and 8 mm since the emission would still be from dust. It is possible that the VLA
could be filtering out this emission, but this emission would not
be filtered in the 1.3 mm and 850 \micron\ images that we presented in Figures 2 \& 3.
Line-contaminated continuum could also
explain these sources and their close association with outflow 
and shock features. \citet{codella2014} had sufficient 
spectral resolution in their continuum bands
to identify line-free regions of the continuum, but the spectral line maps shown in 
Figure 4 of \citet{maury2014} have point-like emission features that appear
coincident with MM2 and MM3 for several molecular lines, notably several methanol and formaldehyde 
transitions.

In summary, MM2 and MM3 have not been detected 
in any other continuum dataset available from a multitude of observatories
spanning wavelengths shorter and longer than the observations of \citet{codella2014}.
Their association with extended envelope emission 
in our 1.3 mm and 850 \micron\ continuum data and multiple line emission 
features in \citet{maury2014} lead us to 
conclude that MM2 and MM3 are most likely not protostellar sources. 
We suggest that the driving source of the north-south outflow 
is VLA1, and the driving source of the east-west outflow is likely VLA2.
Despite the high-resolution of our observations there is still significant blending
of the outflow components and higher resolution and higher sensitivity data will
be needed to definitively disentangle the outflow driving sources.

\subsection{A Compact, Embedded Disk around IRAS2A VLA1?}

The continuum emission of IRAS2A VLA is clearly resolved perpendicular to the outflow
as shown in Figure 1. Furthermore, Gaussian fits to the data all indicate
that the source is resolved and extended perpendicular to the outflow, a possible
indication that we have detected a compact disk surrounding VLA1. The deconvolved size of this
disk-like structure is quite small, just 0\farcs055 ($\sim$13 AU) in diameter. Thus, 
if this is indeed a disk surrounding VLA1, it is about 10 times smaller than the other known Class 0
disks \citep{tobin2012,murillo2013,lindberg2014,codella2014b}. We emphasize
that if we are detecting the protostellar disk in the Ka-band data, we may not 
be capturing its full extent. This could be due to surface brightness sensitivity limits and possibly
the distribution of dust grain sizes within the disk itself. More-evolved
proto-planetary disks are found to be more compact at 8 mm than when viewed at shorter wavelengths due
to grain growth and radial drift of dust grains \citep{perez2012}, and this process can already
be at work in the protostellar phase \citep{birnstiel2010}. Therefore the
actual size of the disk could be larger than the structure we resolve with the VLA. 
However, the CARMA 1.3 mm data presented in Figure 2 and SMA 850 \micron\ data in 
Figure 3 do not have indications of a disk $>$ 100 AU in diameter.

The position angle of the disk-like structure is also in nearly the same direction as the velocity gradient
observed in methanol emission by \citet{maret2014}. However, the methanol
emission was not thought to be associated with the protostellar disk due to its kinematics
being inconsistent with Keplerian rotation. Furthermore, the size of the methanol emitting region
was 0\farcs44 (deconvolved), about 8 times larger than the disk-like structure. \citet{brinch2009}
had also investigated the kinematics of IRAS2A and found no strong evidence of rotation
on $\sim$200 AU scales. Thus, it seems likely that the disk around IRAS2A is smaller than $\sim$200 AU
and perhaps comparable to the size we measure for the disk-like object. Of course, future sub/millimeter 
observations of both dust continuum and molecular lines at high resolution will be crucial to the
further characterization of the disk around this source.

\subsection{Origin of IRAS2A System}

The mechanism by which binaries form remains a subject of debate. 
Current theoretical work suggests two dominant mechanisms for the formation of 
multiple systems: fragmentation on cloud or filament scales 
\citep{bb1993,bonnell1993,padoan2002,offner2010} or on disk scales 
\citep{bonnell1994a,kratter2010,zhu2012}. 
Fragmentation on core scales is likely due to turbulence 
(although early work considered rotation), while fragmentation on 
disk scales is due to gravitational instability.  
Both disk fragmentation and core fragmentation are likely to contribute to
protostellar multiplicity in different environments and for 
different stellar masses. Because of the small projected 
separation ($\sim$143 AU), the IRAS2A system naively seems a good candidate for the 
disk fragmentation scenario. However, the outflows tell a different story. If 
the outflow from each component of the IRAS2A system is perpendicular to the
(inner) disk surrounding each protostar, as expected from the currently favored
magneto-centrifugal mechanism for outflow generation \citep{blandford1982, frank2014},
the disk around the VLA2 should be perpendicular to that 
of the VLA1. The large relative inclination suggests that the two do not share 
the same net angular momentum vector, as would be expected if the stars formed in the same protostellar disk.

The turbulent fragmentation model is somewhat more consistent with the data. 
Turbulent fragmentation typically occurs on larger scales, $\sim$1000 AU, 
however separation evolution down to $\sim$100 AU on timescales of $10^4$ yrs is 
observed in simulations \citep{offner2010}. If the binary underwent substantial
 migration, one might expect to see it reflected in changes to the outflow, which 
we do not observe. The blue-shifted side of the east-west outflow that we associate 
with VLA2 extends $\sim$0.1 pc (90\arcsec). Assuming a typical outflow 
velocity between 10 \kms\ and 100 \kms\ \citep[e.g.,][]{frank2014} the 
dynamical age of the outflow could be between 1000 yr and 10000 yr. This calculation  assumes that
the outflow is exactly in the plane of the sky. Since the outflow is observed closer to edge-on
than face-on \citep{plunkett2013}, the inclination correction would likely be a factor 
of $\sim$2. The relatively constant outflow 
direction over the last 1000 yr to 10000 yr suggests 
that the orbital migration has been smooth and/or slow enough 
to preserve the orientation. Such smooth motion rules out dynamical interaction with 
another body as the means to decrease the separation. The concept of ``orbital evolution'' 
for such systems can be somewhat misleading: shrinking separations can occur because the 
relative velocities are convergent due to global collapse of the birth core or filament. 
Such evolution will not necessarily cause precession of the star$+$disk angular momentum vector.

Alternatively, the strong outflow may have only began to be driven
after much of the orbital evolution to 
its current location had finished.
A system with a combined mass of 1 $M_{\sun}$ and a semi-major
axis of 142 AU would have an orbital period of $\sim$1,700 yr. If this orbit
is highly elliptical, then the bulk of the orbit will be spent at distances
far from the dominant component (assumed to be VLA1), and it is possible that
the apparent outflow direction would not be appreciably changed. 
The small `wiggle' observed in the outflow on scales $<$ 1000 AU could possibly result
from movement of VLA2.

Another possibility, albeit unlikely, is that IRAS2A is simply a superposition of
two sources along the line of sight, and the true separation is much larger than 
the projected separation. 
However, the peak surface density
of young stellar objects in NGC 1333 is $\sim$200 pc$^{-2}$ \citep{gutermuth2008}, 
and the probability of two sources being within
a projected radius of 143 AU is $\sim$10$^{-4}$. 
Moreover, molecular line observations toward IRAS2A
by \citet{volgenau2006} do not show strong evidence for two cores aligned along the line of sight.
The C$^{18}$O, H$^{13}$CO$^+$, and N$_{2}$H$^+$ spectra have single peaks at resolutions
of 10\arcsec\ and less. Thus, the simplest and most likely possibility is that these two sources are 
in close physical proximity. High resolution molecular line observations that resolve
the two sources are necessary to determine their relative velocities.

We propose that VLA2 most likely formed via turbulent fragmentation. 
Although misaligned disks are consistent with more-evolved, Class II, binary systems
\citep{jensen2014,williams2014}, 
we can only infer the disk misalignment based on outflow directions. Thus, IRAS2A and systems
like it could be the predecessors to misaligned proto-planetary disk systems.

\subsection{Outflow Emission Structure}

The outflow emission from IRAS2A has a few features worth further discussion. 
The bi-polar east-west outflow that we suggest originates from VLA2 appears 
quite linear and well-collimated
in the CO ($J=1\rightarrow0$) map in Figure 5. However, on smaller-scales the
outflow is not quite as symmetric in appearance. The CO emission plots in
Figure 6 show a bright lump of red-shifted emission 
which is consistent with the location of MM2
from \citet{codella2014}
as well as the locations of the peak SiO and SO emission 
observed by those authors. However, no equivalent blue-shifted emission was observed on the opposing 
side of the protostar, in fact there is some low-velocity
red-shifted SO on the blue-shifted side of the outflow \citep[][their Figure 1 and Figure B.2]{codella2014}. 
In the case of the CO emission, the blue-shifted emission near VLA2 
could simply be blended with the emission from VLA1. \citet{codella2014}
suggested that the lack of a strong blue-shifted component near the protostellar
source may be due to monopolar jet activity or the jet driving into
an inhomogeneous medium where the blue-shifted side is not readily apparent. This statement
may hold even for the case of VLA2 being the driving source.

The bright lump of CO emission associated with the position of MM2 and the SiO and SO emission
in \citet{codella2014} is also slightly offset from the main axis of the larger-scale flow (see Figure 6).
The offset of the red-shifted emission near the protostar, relative to the larger-scale east-west flow, 
could result from outflow
precession, where the more recent ejection events have a slightly different axis than the
previous episodes. There is some evidence for blue-shifted emission being offset
to the north from the larger-scale east-west flow, but this is offset from the 
protostar along the outflow by $\sim$1000 AU. Alternatively, 
the shift in the outflow could be due to the orbital motion of VLA2 around
VLA1. This is plausible because the time required for an 
outflow of 10 \kms\ - 100 \kms\ to propagate
1000 AU is $\sim$ 50 yr - 500 yr, and the orbital period of VLA2 would be $\sim$1700 yr, 
assuming VLA1 has a mass of 1 $M_{\sun}$ and the semi-major axis is equal to the current separation. 
Thus, during the course of outflow propagation VLA2 could have moved through enough
of its orbit to exhibit a change in direction. This depends on the combined mass of VLA1 and VLA2 and
the orbital parameters.

\section{Conclusions}

We have presented VLA imaging of the NGC 1333 IRAS2A region at 8 mm, 1 cm, 4 cm, and 6.4 cm
with a best resolution of $\sim$0\farcs06 (14 AU), resolving two 
sources toward IRAS2A (VLA1 and VLA2), separated by 0\farcs621 $\pm$ 0.006 ($\sim$143 AU).
Emission at 1.3 mm from VLA2 is tentatively detected in complementary 
CARMA images at the 5$\sigma$ level. 
We do not detect the candidate companion sources MM2 and MM3 identified
by \citet{codella2014} in our VLA data. Moreover,
these sources were also not detected in
sufficiently sensitive CARMA 1.3 mm data and SMA 850 \micron\
data, as well as other PdBI observations. Extended structure
is observed toward the locations of MM2 and MM3 at both 1.3 mm and 850 \micron, an 
indication that these sources could be spatial filtering artifacts or line-contaminated continuum.
$^{12}$CO ($J=2\rightarrow1$) and ($J=3\rightarrow2$) 
observations show that MM2 and MM3 are located right in outflow lobes. We conclude that VLA1
is the driving source of the north-south outflow and 
propose that VLA2 is the driving source of the east-west outflow.  

VLA1 is found to be extended perpendicular to the direction of the outflow in the Ka-band imaging.
The deconvolved size of the major axis is 0\farcs055 ($\sim$13 AU). This structure could be part of the
protostellar disk around VLA1. In contrast, VLA2 has no evidence of extended structure and
is consistent with being a point source.

The spectral indices at centimeter wavelengths
for both VLA1 and VLA2 are consistent with moderately to very optically thick free-free emission. 
The free-free emission from VLA1
accounts for about half the total flux at 8 mm, the other half from thermal dust 
emission. VLA2, on the other hand, appears dominated by free-free emission until a wavelength of $\sim$1 mm, 
indicative of very little circumstellar dust.

The orthogonal outflows from the sources are strong evidence against disk fragmentation being the formation
route for the VLA1 and VLA2 binary system. Turbulent fragmentation with orbital evolution is possible and most likely; 
however, the outflow shows no large-scale changes in direction 
over the past 1000 yr - 10000 yr. Future kinematic observations at $\sim$0\farcs25 resolution have the promise
to better characterize the origins of this proto-binary system.

The authors wish to thank the anonymous referee for a constructive report that
improve the paper.
We also wish to thank L. Kristensen and M. Persson for stimulating discussions 
regarding this work. J.J.T. acknowledges support provided by NASA through Hubble Fellowship 
grant \#HST-HF-51300.01-A awarded by the Space Telescope Science Institute, which is 
operated by the Association of Universities for Research in Astronomy, 
Inc., for NASA, under contract NAS 5-26555. 
M.M.D. acknowledges support from the Submillimeter Array through an SMA 
postdoctoral fellowship.  
L.W.L. acknowledges support from the Laboratory for Astronomical 
Imaging at the University of Illinois and the NSF under grant AST-07-09206.
D.S.-C. acknowledges support provided by the NSF through award GSSP 2013-06713
from the NRAO. C.M. acknowledges financial support
from the U.S. National Science Foundation through award AST-1313428.
S.I.S. acknowledges support from the National Science and Engineering Research
Council of Canada (NSERC) Postdoctoral Fellowship.
Z.Y.L. is supported in part by NNX14AB38G and AST-1313083.
A.L.P. is supported by the National Science Foundation 
Graduate Research Fellowship under Grant No. DGE-1122492.
L.M.P acknowledges support from the Jansky Fellowship program of the National
Radio Astronomy Observatory.
Support for CARMA construction was derived from the states of Illinois, California, and Maryland, 
the James S. McDonnell Foundation, the Gordon and Betty Moore Foundation, the Kenneth T. and 
Eileen L. Norris Foundation, the University of Chicago, the Associates of the California 
Institute of Technology, and the National Science Foundation. Ongoing CARMA development 
and operations are supported by the National Science Foundation under a cooperative 
agreement, and by the CARMA partner universities. The Submillimeter Array is a joint 
project between the Smithsonian Astrophysical Observatory and the Academia Sinica 
Institute of Astronomy and Astrophysics and is funded by the Smithsonian 
Institution and the Academia Sinica. The National Radio Astronomy 
Observatory is a facility of the National Science Foundation 
operated under cooperative agreement by Associated Universities, Inc.

The authors wish to recognize and acknowledge the very significant 
cultural role and reverence that the summit of Mauna Kea has always 
had within the indigenous Hawaiian community.  We are most fortunate 
to have the opportunity to conduct observations from this mountain. 

{\it Facilities:}  \facility{VLA}, \facility{CARMA}, \facility{SMA}, \facility{\textit{Spitzer}}

\begin{small}
\bibliographystyle{apj}
\bibliography{ms}

\begin{thebibliography}{}
\expandafter\ifx\csname natexlab\endcsname\relax\def\natexlab#1{#1}\fi

\bibitem[{{Andre} {et~al.}(1993){Andre}, {Ward-Thompson}, \&
  {Barsony}}]{andre1993}
{Andre}, P., {Ward-Thompson}, D., \& {Barsony}, M. 1993, \apj, 406, 122

\bibitem[{{Anglada}(1995)}]{anglada1995}
{Anglada}, G. 1995, in Revista Mexicana de Astronomia y Astrofisica, vol. 27,
  Vol.~1, Revista Mexicana de Astronomia y Astrofisica Conference Series, ed.
  S.~{Lizano} \& J.~M. {Torrelles}, 67

\bibitem[{{Anglada} {et~al.}(1998){Anglada}, {Villuendas}, {Estalella},
  {Beltr{\'a}n}, {Rodr{\'{\i}}guez}, {Torrelles}, \& {Curiel}}]{anglada1998}
{Anglada}, G., {Villuendas}, E., {Estalella}, R., {et~al.} 1998, \aj, 116, 2953

\bibitem[{{Birnstiel} {et~al.}(2010){Birnstiel}, {Dullemond}, \&
  {Brauer}}]{birnstiel2010}
{Birnstiel}, T., {Dullemond}, C.~P., \& {Brauer}, F. 2010, \aap, 513, A79

\bibitem[{{Blandford} \& {Payne}(1982)}]{blandford1982}
{Blandford}, R.~D., \& {Payne}, D.~G. 1982, \mnras, 199, 883

\bibitem[{{Bonnell} \& {Bastien}(1993)}]{bonnell1993}
{Bonnell}, I., \& {Bastien}, P. 1993, \apj, 406, 614

\bibitem[{{Bonnell} \& {Bate}(1994{\natexlab{a}})}]{bonnell1994a}
{Bonnell}, I.~A., \& {Bate}, M.~R. 1994{\natexlab{a}}, \mnras, 269, L45

\bibitem[{{Bonnell} \& {Bate}(1994{\natexlab{b}})}]{bonnell1994b}
---. 1994{\natexlab{b}}, \mnras, 271, 999

\bibitem[{{Bontemps} {et~al.}(1996){Bontemps}, {Andre}, {Terebey}, \&
  {Cabrit}}]{bontemps1996}
{Bontemps}, S., {Andre}, P., {Terebey}, S., \& {Cabrit}, S. 1996, \aap, 311,
  858

\bibitem[{{Boss} \& {Keiser}(2013)}]{boss2013}
{Boss}, A.~P., \& {Keiser}, S.~A. 2013, \apj, 764, 136

\bibitem[{{Boss} \& {Keiser}(2014)}]{boss2014}
---. 2014, ArXiv e-prints, arXiv:1408.2479

\bibitem[{{Brinch} {et~al.}(2009){Brinch}, {J{\o}rgensen}, \&
  {Hogerheijde}}]{brinch2009}
{Brinch}, C., {J{\o}rgensen}, J.~K., \& {Hogerheijde}, M.~R. 2009, \aap, 502,
  199

\bibitem[{{Burkert} \& {Bodenheimer}(1993)}]{bb1993}
{Burkert}, A., \& {Bodenheimer}, P. 1993, \mnras, 264, 798

\bibitem[{{Carrasco-Gonz{\'a}lez} {et~al.}(2008){Carrasco-Gonz{\'a}lez},
  {Anglada}, {Rodr{\'{\i}}guez}, {Torrelles}, {Osorio}, \&
  {Girart}}]{carrasco2008}
{Carrasco-Gonz{\'a}lez}, C., {Anglada}, G., {Rodr{\'{\i}}guez}, L.~F., {et~al.}
  2008, \apj, 676, 1073

\bibitem[{{Cassen} \& {Moosman}(1981)}]{cassen1981}
{Cassen}, P., \& {Moosman}, A. 1981, \icarus, 48, 353

\bibitem[{{Chen} {et~al.}(2010){Chen}, {Arce}, {Zhang}, {Bourke}, {Launhardt},
  {Schmalzl}, \& {Henning}}]{chen2010}
{Chen}, X., {Arce}, H.~G., {Zhang}, Q., {et~al.} 2010, \apj, 715, 1344

\bibitem[{{Chen} {et~al.}(2013){Chen}, {Arce}, {Zhang}, {Bourke}, {Launhardt},
  {J{\o}rgensen}, {Lee}, {Foster}, {Dunham}, {Pineda}, \& {Henning}}]{chen2013}
---. 2013, \apj, 768, 110

\bibitem[{{Codella} {et~al.}(2014{\natexlab{a}}){Codella}, {Maury}, {Gueth},
  {Maret}, {Belloche}, {Cabrit}, \& {Andr{\'e}}}]{codella2014}
{Codella}, C., {Maury}, A.~J., {Gueth}, F., {et~al.} 2014{\natexlab{a}}, \aap,
  563, L3

\bibitem[{{Codella} {et~al.}(2014{\natexlab{b}}){Codella}, {Cabrit}, {Gueth},
  {Podio}, {Leurini}, {Bachiller}, {Gusdorf}, {Lefloch}, {Nisini}, {Tafalla},
  \& {Yvart}}]{codella2014b}
{Codella}, C., {Cabrit}, S., {Gueth}, F., {et~al.} 2014{\natexlab{b}}, \aap,
  568, L5

\bibitem[{{Connelley} {et~al.}(2008){Connelley}, {Reipurth}, \&
  {Tokunaga}}]{connelley2008}
{Connelley}, M.~S., {Reipurth}, B., \& {Tokunaga}, A.~T. 2008, \aj, 135, 2526

\bibitem[{{Coutens} {et~al.}(2014){Coutens}, {J{\o}rgensen}, {Persson}, {van
  Dishoeck}, {Vastel}, \& {Taquet}}]{coutens2014}
{Coutens}, A., {J{\o}rgensen}, J.~K., {Persson}, M.~V., {et~al.} 2014, \apjl,
  792, L5

\bibitem[{{Curiel} {et~al.}(1989){Curiel}, {Rodriguez}, {Bohigas}, {Roth},
  {Canto}, \& {Torrelles}}]{curiel1989}
{Curiel}, S., {Rodriguez}, L.~F., {Bohigas}, J., {et~al.} 1989, Astrophysical
  Letters and Communications, 27, 299

\bibitem[{{Draine} {et~al.}(1983){Draine}, {Roberge}, \&
  {Dalgarno}}]{draine1983}
{Draine}, B.~T., {Roberge}, W.~G., \& {Dalgarno}, A. 1983, \apj, 264, 485

\bibitem[{{Duch{\^e}ne} \& {Kraus}(2013)}]{duchene2013}
{Duch{\^e}ne}, G., \& {Kraus}, A. 2013, \araa, 51, 269

\bibitem[{{Engargiola} \& {Plambeck}(1999)}]{engarg1999}
{Engargiola}, G., \& {Plambeck}, R.~L. 1999, in The Physics and Chemistry of
  the Interstellar Medium, ed. V.~{Ossenkopf}, J.~{Stutzki}, \&
  G.~{Winnewisser}, 291

\bibitem[{{Enoch} {et~al.}(2009){Enoch}, {Evans}, {Sargent}, \&
  {Glenn}}]{enoch2009}
{Enoch}, M.~L., {Evans}, N.~J., {Sargent}, A.~I., \& {Glenn}, J. 2009, \apj,
  692, 973

\bibitem[{{Enoch} {et~al.}(2010){Enoch}, {Lee}, {Harvey}, {Dunham}, \&
  {Schnee}}]{enoch2010}
{Enoch}, M.~L., {Lee}, J.-E., {Harvey}, P., {Dunham}, M.~M., \& {Schnee}, S.
  2010, \apjl, 722, L33

\bibitem[{{Frank} {et~al.}(2014){Frank}, {Ray}, {Cabrit}, {Hartigan}, {Arce},
  {Bacciotti}, {Bally}, {Benisty}, {Eisl{\"o}ffel}, {G{\"u}del}, {Lebedev},
  {Nisini}, \& {Raga}}]{frank2014}
{Frank}, A., {Ray}, T.~P., {Cabrit}, S., {et~al.} 2014, ArXiv e-prints,
  arXiv:1402.3553

\bibitem[{{Girart} {et~al.}(2009){Girart}, {Rao}, \& {Estalella}}]{girart2009}
{Girart}, J.~M., {Rao}, R., \& {Estalella}, R. 2009, \apj, 694, 56

\bibitem[{{Gutermuth} {et~al.}(2008){Gutermuth}, {Myers}, {Megeath}, {Allen},
  {Pipher}, {Muzerolle}, {Porras}, {Winston}, \& {Fazio}}]{gutermuth2008}
{Gutermuth}, R.~A., {Myers}, P.~C., {Megeath}, S.~T., {et~al.} 2008, \apj, 674,
  336

\bibitem[{{Hirano} {et~al.}(1999){Hirano}, {Kamazaki}, {Mikami}, {Ohashi}, \&
  {Umemoto}}]{hirano1999}
{Hirano}, N., {Kamazaki}, T., {Mikami}, H., {Ohashi}, N., \& {Umemoto}, T.
  1999, in Star Formation 1999, ed. T.~{Nakamoto}, 181--182

\bibitem[{{Hirota} {et~al.}(2011){Hirota}, {Honma}, {Imai}, {Sunada}, {Ueno},
  {Kobayashi}, \& {Kawaguchi}}]{hirota2011}
{Hirota}, T., {Honma}, M., {Imai}, H., {et~al.} 2011, \pasj, 63, 1

\bibitem[{{Hirota} {et~al.}(2008){Hirota}, {Bushimata}, {Choi}, {Honma},
  {Imai}, {Iwadate}, {Jike}, {Kameya}, {Kamohara}, {Kan-Ya}, {Kawaguchi},
  {Kijima}, {Kobayashi}, {Kuji}, {Kurayama}, {Manabe}, {Miyaji}, {Nagayama},
  {Nakagawa}, {Oh}, {Omodaka}, {Oyama}, {Sakai}, {Sasao}, {Sato}, {Shibata},
  {Tamura}, \& {Yamashita}}]{hirota2008}
{Hirota}, T., {Bushimata}, T., {Choi}, Y.~K., {et~al.} 2008, \pasj, 60, 37

\bibitem[{{Ho} {et~al.}(2004){Ho}, {Moran}, \& {Lo}}]{ho2004}
{Ho}, P.~T.~P., {Moran}, J.~M., \& {Lo}, K.~Y. 2004, \apjl, 616, L1

\bibitem[{{Hollenbach} \& {McKee}(1989)}]{hollenbach1989}
{Hollenbach}, D., \& {McKee}, C.~F. 1989, \apj, 342, 306

\bibitem[{{Hull} {et~al.}(2014){Hull}, {Plambeck}, {Kwon}, {Bower},
  {Carpenter}, {Crutcher}, {Fiege}, {Franzmann}, {Hakobian}, {Heiles}, {Houde},
  {Hughes}, {Lamb}, {Looney}, {Marrone}, {Matthews}, {Pillai}, {Pound},
  {Rahman}, {Sandell}, {Stephens}, {Tobin}, {Vaillancourt}, {Volgenau}, \&
  {Wright}}]{hull2014}
{Hull}, C.~L.~H., {Plambeck}, R.~L., {Kwon}, W., {et~al.} 2014, \apjs, 213, 13

\bibitem[{{Inutsuka} \& {Miyama}(1992)}]{inutsuka1992}
{Inutsuka}, S.-I., \& {Miyama}, S.~M. 1992, \apj, 388, 392

\bibitem[{{Jensen} \& {Akeson}(2014)}]{jensen2014}
{Jensen}, E.~L.~N., \& {Akeson}, R. 2014, \nat, 511, 567

\bibitem[{{J{\o}rgensen} {et~al.}(2005){J{\o}rgensen}, {Bourke}, {Myers},
  {Sch{\"o}ier}, {van Dishoeck}, \& {Wilner}}]{jorgensen2005}
{J{\o}rgensen}, J.~K., {Bourke}, T.~L., {Myers}, P.~C., {et~al.} 2005, \apj,
  632, 973

\bibitem[{{J{\o}rgensen} {et~al.}(2004{\natexlab{a}}){J{\o}rgensen},
  {Hogerheijde}, {Blake}, {van Dishoeck}, {Mundy}, \&
  {Sch{\"o}ier}}]{jorgensen2004b}
{J{\o}rgensen}, J.~K., {Hogerheijde}, M.~R., {Blake}, G.~A., {et~al.}
  2004{\natexlab{a}}, \aap, 415, 1021

\bibitem[{{J{\o}rgensen} {et~al.}(2004{\natexlab{b}}){J{\o}rgensen},
  {Hogerheijde}, {van Dishoeck}, {Blake}, \& {Sch{\"o}ier}}]{jorgensen2004a}
{J{\o}rgensen}, J.~K., {Hogerheijde}, M.~R., {van Dishoeck}, E.~F., {Blake},
  G.~A., \& {Sch{\"o}ier}, F.~L. 2004{\natexlab{b}}, \aap, 413, 993

\bibitem[{{J{\o}rgensen} {et~al.}(2006){J{\o}rgensen}, {Harvey}, {Evans},
  {Huard}, {Allen}, {Porras}, {Blake}, {Bourke}, {Chapman}, {Cieza}, {Koerner},
  {Lai}, {Mundy}, {Myers}, {Padgett}, {Rebull}, {Sargent}, {Spiesman},
  {Stapelfeldt}, {van Dishoeck}, {Wahhaj}, \& {Young}}]{jorgensen2006}
{J{\o}rgensen}, J.~K., {Harvey}, P.~M., {Evans}, II, N.~J., {et~al.} 2006,
  \apj, 645, 1246

\bibitem[{{Kratter} {et~al.}(2010){Kratter}, {Matzner}, {Krumholz}, \&
  {Klein}}]{kratter2010}
{Kratter}, K.~M., {Matzner}, C.~D., {Krumholz}, M.~R., \& {Klein}, R.~I. 2010,
  \apj, 708, 1585

\bibitem[{{Kraus} {et~al.}(2011){Kraus}, {Ireland}, {Martinache}, \&
  {Hillenbrand}}]{kraus2011}
{Kraus}, A.~L., {Ireland}, M.~J., {Martinache}, F., \& {Hillenbrand}, L.~A.
  2011, \apj, 731, 8

\bibitem[{{Lada}(1987)}]{lada1987}
{Lada}, C.~J. 1987, in IAU Symp. 115: Star Forming Regions, ed. M.~{Peimbert}
  \& J.~{Jugaku}, 1--17

\bibitem[{{Lada}(2006)}]{lada2006}
{Lada}, C.~J. 2006, \apjl, 640, L63

\bibitem[{{Lindberg} {et~al.}(2014){Lindberg}, {J{\o}rgensen}, {Brinch},
  {Haugb{\o}lle}, {Bergin}, {Harsono}, {Persson}, {Visser}, \&
  {Yamamoto}}]{lindberg2014}
{Lindberg}, J.~E., {J{\o}rgensen}, J.~K., {Brinch}, C., {et~al.} 2014, \aap,
  566, A74

\bibitem[{{Looney} {et~al.}(2000){Looney}, {Mundy}, \& {Welch}}]{looney2000}
{Looney}, L.~W., {Mundy}, L.~G., \& {Welch}, W.~J. 2000, \apj, 529, 477

\bibitem[{{Maret} {et~al.}(2014){Maret}, {Belloche}, {Maury}, {Gueth},
  {Andr{\'e}}, {Cabrit}, {Codella}, \& {Bontemps}}]{maret2014}
{Maret}, S., {Belloche}, A., {Maury}, A.~J., {et~al.} 2014, \aap, 563, L1

\bibitem[{{Markwardt}(2009)}]{markwardt2009}
{Markwardt}, C.~B. 2009, in Astronomical Society of the Pacific Conference
  Series, Vol. 411, Astronomical Data Analysis Software and Systems XVIII, ed.
  {D.~A.~Bohlender, D.~Durand, \& P.~Dowler}, 251

\bibitem[{{Maury} {et~al.}(2010){Maury}, {Andr{\'e}}, {Hennebelle}, {Motte},
  {Stamatellos}, {Bate}, {Belloche}, {Duch{\^e}ne}, \& {Whitworth}}]{maury2010}
{Maury}, A.~J., {Andr{\'e}}, P., {Hennebelle}, P., {et~al.} 2010, \aap, 512,
  A40

\bibitem[{{Maury} {et~al.}(2014){Maury}, {Belloche}, {Andr{\'e}}, {Maret},
  {Gueth}, {Codella}, {Cabrit}, {Testi}, \& {Bontemps}}]{maury2014}
{Maury}, A.~J., {Belloche}, A., {Andr{\'e}}, P., {et~al.} 2014, \aap, 563, L2

\bibitem[{{Melis} {et~al.}(2011){Melis}, {Duch{\^e}ne}, {Chomiuk}, {Palmer},
  {Perrin}, {Maddison}, {M{\'e}nard}, {Stapelfeldt}, {Pinte}, \&
  {Duvert}}]{melis2011}
{Melis}, C., {Duch{\^e}ne}, G., {Chomiuk}, L., {et~al.} 2011, \apjl, 739, L7

\bibitem[{{Murillo} \& {Lai}(2013)}]{murillo2013}
{Murillo}, N.~M., \& {Lai}, S.-P. 2013, \apjl, 764, L15

\bibitem[{{Offner} {et~al.}(2010){Offner}, {Kratter}, {Matzner}, {Krumholz}, \&
  {Klein}}]{offner2010}
{Offner}, S.~S.~R., {Kratter}, K.~M., {Matzner}, C.~D., {Krumholz}, M.~R., \&
  {Klein}, R.~I. 2010, \apj, 725, 1485

\bibitem[{{Padoan} \& {Nordlund}(2002)}]{padoan2002}
{Padoan}, P., \& {Nordlund}, {\AA}. 2002, \apj, 576, 870

\bibitem[{{P{\'e}rez} {et~al.}(2012){P{\'e}rez}, {Carpenter}, {Chandler},
  {Isella}, {Andrews}, {Ricci}, {Calvet}, {Corder}, {Deller}, {Dullemond},
  {Greaves}, {Harris}, {Henning}, {Kwon}, {Lazio}, {Linz}, {Mundy}, {Sargent},
  {Storm}, {Testi}, \& {Wilner}}]{perez2012}
{P{\'e}rez}, L.~M., {Carpenter}, J.~M., {Chandler}, C.~J., {et~al.} 2012,
  \apjl, 760, L17

\bibitem[{{Persson} {et~al.}(2012){Persson}, {J{\o}rgensen}, \& {van
  Dishoeck}}]{persson2012}
{Persson}, M.~V., {J{\o}rgensen}, J.~K., \& {van Dishoeck}, E.~F. 2012, \aap,
  541, A39

\bibitem[{{Pezzuto} {et~al.}(2012){Pezzuto}, {Elia}, {Schisano}, {Strafella},
  {Di Francesco}, {Sadavoy}, {Andr{\'e}}, {Benedettini}, {Bernard}, {di
  Giorgio}, {Facchini}, {Hennemann}, {Hill}, {K{\"o}nyves}, {Molinari},
  {Motte}, {Nguyen-Luong}, {Peretto}, {Pestalozzi}, {Polychroni}, {Rygl},
  {Saraceno}, {Schneider}, {Spinoglio}, {Testi}, {Ward-Thompson}, \&
  {White}}]{pezzuto2012}
{Pezzuto}, S., {Elia}, D., {Schisano}, E., {et~al.} 2012, \aap, 547, A54

\bibitem[{{Pineda} {et~al.}(2011){Pineda}, {Arce}, {Schnee}, {Goodman},
  {Bourke}, {Foster}, {Robitaille}, {Tanner}, {Kauffmann}, {Tafalla},
  {Caselli}, \& {Anglada}}]{pineda2011}
{Pineda}, J.~E., {Arce}, H.~G., {Schnee}, S., {et~al.} 2011, \apj, 743, 201

\bibitem[{{Plunkett} {et~al.}(2013){Plunkett}, {Arce}, {Corder}, {Mardones},
  {Sargent}, \& {Schnee}}]{plunkett2013}
{Plunkett}, A.~L., {Arce}, H.~G., {Corder}, S.~A., {et~al.} 2013, \apj, 774, 22

\bibitem[{{Raghavan} {et~al.}(2010){Raghavan}, {McAlister}, {Henry}, {Latham},
  {Marcy}, {Mason}, {Gies}, {White}, \& {ten Brummelaar}}]{raghavan2010}
{Raghavan}, D., {McAlister}, H.~A., {Henry}, T.~J., {et~al.} 2010, \apjs, 190,
  1

\bibitem[{{Reipurth} {et~al.}(2014){Reipurth}, {Clarke}, {Boss}, {Goodwin},
  {Rodriguez}, {Stassun}, {Tokovinin}, \& {Zinnecker}}]{reipurth2014}
{Reipurth}, B., {Clarke}, C.~J., {Boss}, A.~P., {et~al.} 2014, ArXiv e-prints,
  arXiv:1403.1907

\bibitem[{{Reipurth} {et~al.}(2002){Reipurth}, {Rodr{\'{\i}}guez}, {Anglada},
  \& {Bally}}]{reipurth2002}
{Reipurth}, B., {Rodr{\'{\i}}guez}, L.~F., {Anglada}, G., \& {Bally}, J. 2002,
  \aj, 124, 1045

\bibitem[{{Rodmann} {et~al.}(2006){Rodmann}, {Henning}, {Chandler}, {Mundy}, \&
  {Wilner}}]{rodmann2006}
{Rodmann}, J., {Henning}, T., {Chandler}, C.~J., {Mundy}, L.~G., \& {Wilner},
  D.~J. 2006, \aap, 446, 211

\bibitem[{{Rodr{\'{\i}}guez} {et~al.}(1999){Rodr{\'{\i}}guez}, {Anglada}, \&
  {Curiel}}]{rodriguez1999}
{Rodr{\'{\i}}guez}, L.~F., {Anglada}, G., \& {Curiel}, S. 1999, \apjs, 125, 427

\bibitem[{{Sadavoy} {et~al.}(2014){Sadavoy}, {Di Francesco}, {Andr{\'e}},
  {Pezzuto}, {Bernard}, {Maury}, {Men'shchikov}, {Motte},
  {Nguy{\tilde}{\^e}n-Lu'o'ng}, {Schneider}, {Arzoumanian}, {Benedettini},
  {Bontemps}, {Elia}, {Hennemann}, {Hill}, {K{\"o}nyves}, {Louvet}, {Peretto},
  {Roy}, \& {White}}]{sadavoy2014}
{Sadavoy}, S.~I., {Di Francesco}, J., {Andr{\'e}}, P., {et~al.} 2014, \apjl,
  787, L18

\bibitem[{{Sana} \& {Evans}(2011)}]{sana2011}
{Sana}, H., \& {Evans}, C.~J. 2011, in IAU Symposium, Vol. 272, IAU Symposium,
  ed. C.~{Neiner}, G.~{Wade}, G.~{Meynet}, \& G.~{Peters}, 474--485

\bibitem[{{Sandell} \& {Knee}(2001)}]{sandellknee2001}
{Sandell}, G., \& {Knee}, L.~B.~G. 2001, \apjl, 546, L49

\bibitem[{{Sandell} {et~al.}(1994){Sandell}, {Knee}, {Aspin}, {Robson}, \&
  {Russell}}]{sandell1994}
{Sandell}, G., {Knee}, L.~B.~G., {Aspin}, C., {Robson}, I.~E., \& {Russell},
  A.~P.~G. 1994, \aap, 285, L1

\bibitem[{{Sault} {et~al.}(1995){Sault}, {Teuben}, \& {Wright}}]{sault1995}
{Sault}, R.~J., {Teuben}, P.~J., \& {Wright}, M.~C.~H. 1995, in Astronomical
  Society of the Pacific Conference Series, Vol.~77, Astronomical Data Analysis
  Software and Systems IV, ed. {R.~A.~Shaw, H.~E.~Payne, \& J.~J.~E.~Hayes},
  433

\bibitem[{{Schnee} {et~al.}(2012){Schnee}, {Sadavoy}, {Di Francesco},
  {Johnstone}, \& {Wei}}]{schnee2012}
{Schnee}, S., {Sadavoy}, S., {Di Francesco}, J., {Johnstone}, D., \& {Wei}, L.
  2012, \apj, 755, 178

\bibitem[{{Shirley} {et~al.}(2007){Shirley}, {Claussen}, {Bourke}, {Young}, \&
  {Blake}}]{shirley2007}
{Shirley}, Y.~L., {Claussen}, M.~J., {Bourke}, T.~L., {Young}, C.~H., \&
  {Blake}, G.~A. 2007, \apj, 667, 329

\bibitem[{{Tobin} {et~al.}(2012){Tobin}, {Hartmann}, {Chiang}, {Wilner},
  {Looney}, {Loinard}, {Calvet}, \& {D'Alessio}}]{tobin2012}
{Tobin}, J.~J., {Hartmann}, L., {Chiang}, H.-F., {et~al.} 2012, \nat, 492, 83

\bibitem[{{Tobin} {et~al.}(2013){Tobin}, {Bergin}, {Hartmann}, {Lee}, {Maret},
  {Myers}, {Looney}, {Chiang}, \& {Friesen}}]{tobin2013}
{Tobin}, J.~J., {Bergin}, E.~A., {Hartmann}, L., {et~al.} 2013, \apj, 765, 18

\bibitem[{{Volgenau} {et~al.}(2006){Volgenau}, {Mundy}, {Looney}, \&
  {Welch}}]{volgenau2006}
{Volgenau}, N.~H., {Mundy}, L.~G., {Looney}, L.~W., \& {Welch}, W.~J. 2006,
  \apj, 651, 301

\bibitem[{{Vorobyov} \& {Basu}(2010)}]{vorobyov2010a}
{Vorobyov}, E.~I., \& {Basu}, S. 2010, \apj, 719, 1896

\bibitem[{{Wakelam} {et~al.}(2005){Wakelam}, {Ceccarelli}, {Castets},
  {Lefloch}, {Loinard}, {Faure}, {Schneider}, \& {Benayoun}}]{wakelam2005}
{Wakelam}, V., {Ceccarelli}, C., {Castets}, A., {et~al.} 2005, \aap, 437, 149

\bibitem[{{Walker-Smith} {et~al.}(2014){Walker-Smith}, {Richer}, {Buckle},
  {Hatchell}, \& {Drabek-Maunder}}]{walkersmith2014}
{Walker-Smith}, S.~L., {Richer}, J.~S., {Buckle}, J.~V., {Hatchell}, J., \&
  {Drabek-Maunder}, E. 2014, \mnras, 440, 3568

\bibitem[{{Williams} {et~al.}(2014){Williams}, {Mann}, {Di Francesco},
  {Johnstone}, {Andrews}, {Bally}, {Ricci}, {Hughes}, \&
  {Matthews}}]{williams2014}
{Williams}, J.~P., {Mann}, R.~K., {Di Francesco}, J., {et~al.} 2014, \apj,
  arXiv:1410.3570

\bibitem[{{Wu} {et~al.}(2004){Wu}, {Wei}, {Zhao}, {Shi}, {Yu}, {Qin}, \&
  {Huang}}]{wu2004}
{Wu}, Y., {Wei}, Y., {Zhao}, M., {et~al.} 2004, \aap, 426, 503

\bibitem[{{Zhu} {et~al.}(2012){Zhu}, {Hartmann}, {Nelson}, \&
  {Gammie}}]{zhu2012}
{Zhu}, Z., {Hartmann}, L., {Nelson}, R.~P., \& {Gammie}, C.~F. 2012, \apj, 746,
  110

\end{thebibliography}
\end{small}

\clearpage

\begin{figure}[!ht]
\begin{center}
\vspace{-20mm}
\includegraphics[trim=0.5cm 4.5cm 0.5cm 4.5cm,scale=0.6]{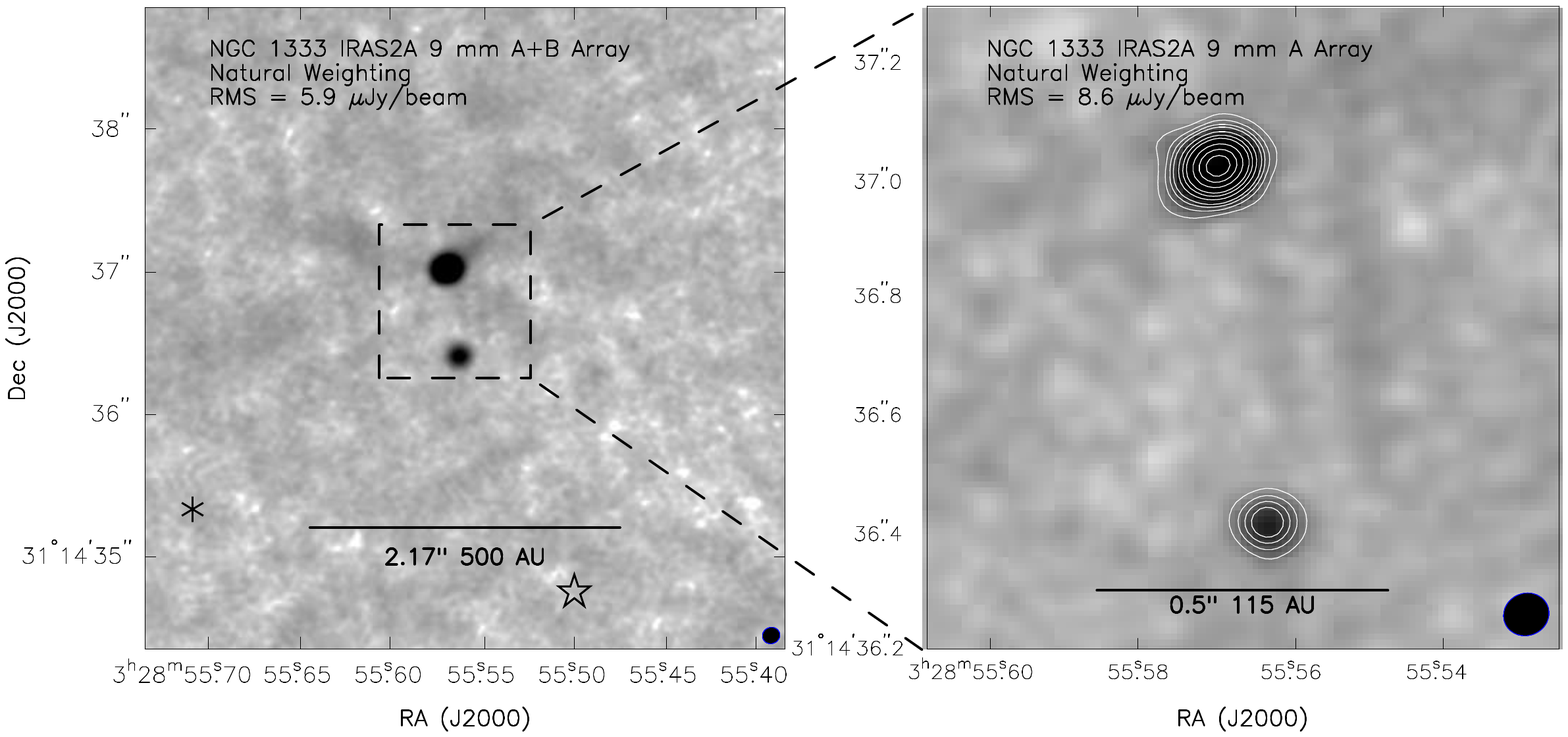}
\includegraphics[trim=0.5cm 4.5cm 0.5cm 4.5cm,scale=0.6]{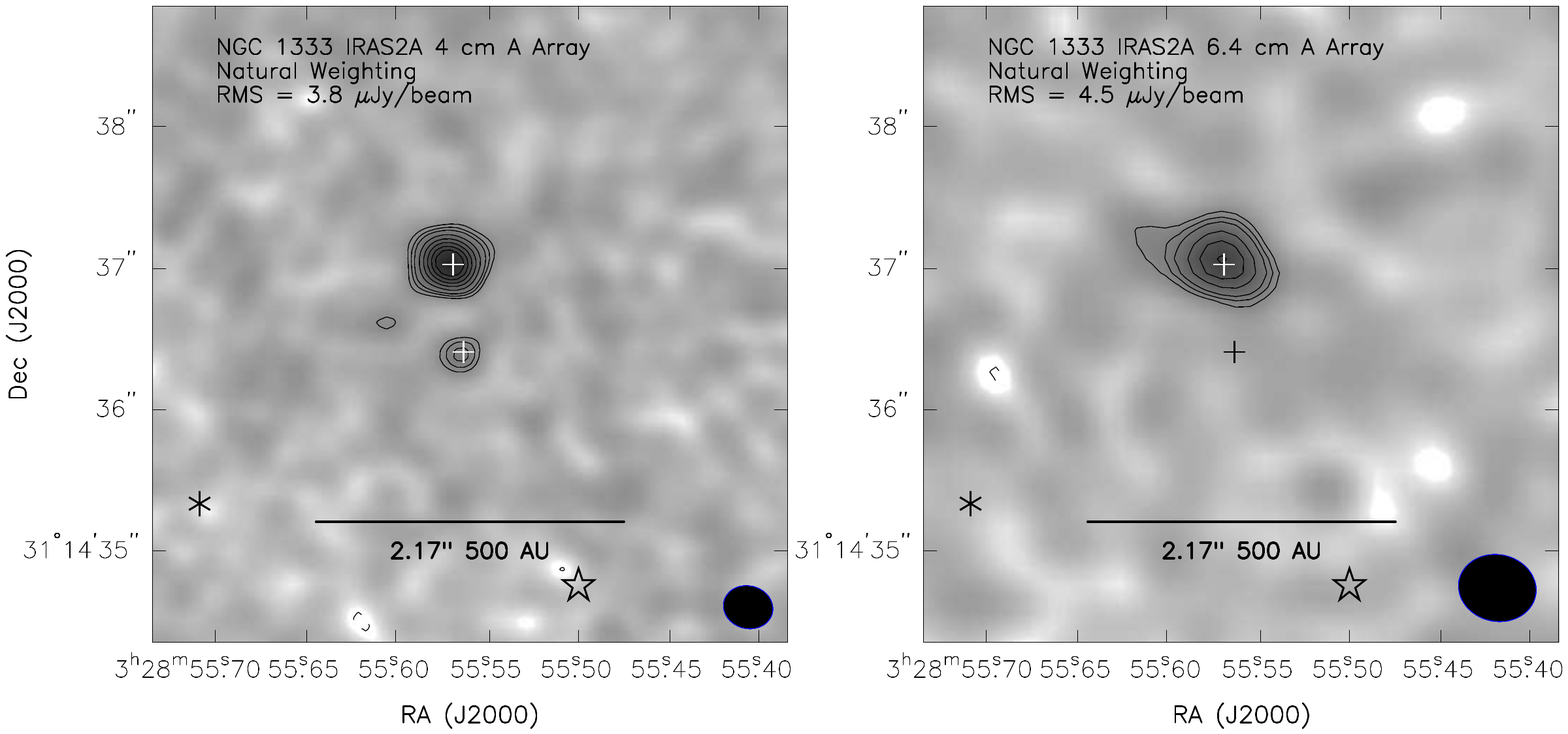}
\end{center}
\caption{NGC 1333 IRAS2A system as observed by the VLA at 9 mm (top panels) 
and 4 cm and 6.4 cm (bottom panels). The top left panel shows the
9mm image from A and B configurations combined, while the zoom-in shown 
in the right panel is from A-array only giving the highest resolution.
The two sources separated by 0\farcs621$\pm$0\farcs006 ($\sim$143 AU) are
clearly resolved in the 9 mm images; in the A+B combined 9 mm image the brighter source
(IRAS2A VLA1) has a signal-to-noise ratio (SNR) of 206 and the 
secondary (IRAS2A VLA2) has a SNR = 41. The sources
are fainter at 4 cm, with VLA1 and VLA2 only having a SNR of 18.5 and 5.4, respectively.
The source VLA1 appears resolved perpendicular to the outflow direction, possibly tracing
a disk. The asterisk and star symbols mark
the positions of MM2 and MM3 respectively \citep{codella2014}. These sources are not
detected in our VLA data. The beam at 9 mm is 0\farcs12 $\times$ 0\farcs11 and 0\farcs08 $\times$ 0\farcs07 for the
A+B and A-only images, respectively. The beam at 4 cm and 6.4 cm is 0\farcs35 $\times$ 0\farcs3
and 0\farcs55 $\times$ 0\farcs47, respectively.
The contour levels in the A-configuration 9 mm image are [-6, 6, 9, 12, 15, 20, 30, ...] $\times$ $\sigma$
and $\sigma$ = 8.6 $\mu$Jy beam$^{-1}$.
The contour levels in the 4 cm and 6.4 cm images are [-3, 3, 4, 5, 7, 9, 11, 13, 15] 
$\times$ $\sigma$ and $\sigma$ = 3.8 $\mu$Jy beam$^{-1}$ and 4.5 $\mu$Jy at 4 cm and 6.4 cm
respectively. The crosses in the lower panels mark the positions of VLA1 and VLA2 derived from the 9 mm
images in the top panels.
}
\label{IRAS2A-VLA}
\end{figure}

\begin{figure}[!ht]
\begin{center}
\includegraphics[scale=0.6]{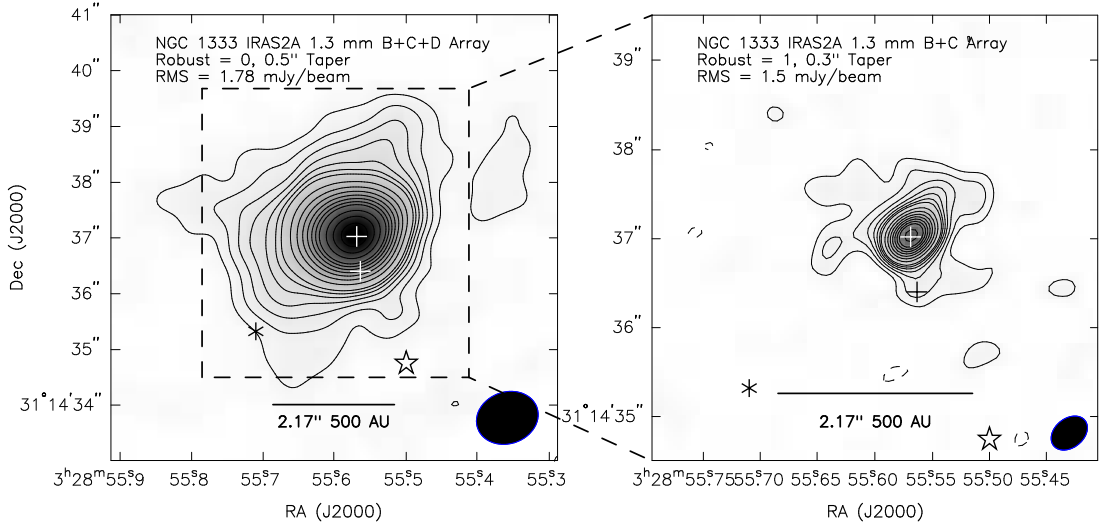}
\end{center}
\caption{NGC 1333 IRAS2A system as observed by CARMA at 1.3 mm. A lower resolution image reconstructed from
a combination of B, C, and D configuration data with Robust = 0 weighting and 0.5\arcsec\ tapering is shown
in the left panel. A higher resolution image is shown in the right panel, combining B and C configuration
data with Robust = 1 weighting and 0.3\arcsec\ tapering.
The crosses mark the positions of VLA1 and VLA2 derived from the 9 mm
images in Figure 1. The 5$\sigma$ contour in the right panel encompasses
the location of VLA2.
The asterisk and star symbols mark
the positions of MM2 and MM3 respectively \citep{codella2014}. The images do not show evidence of
emission at the locations of MM2 and MM3; 
MM2/MM3 should have been detected with $\sim$8$\sigma$/11.8$\sigma$ in the lower 
resolution image and with greater SNR in the
higher resolution image. The contour levels are in the left and right panels are [-3, 3, 5, 7, 9, 11, 13,\
15, 20, 25, 30, 35, 40, 45, 50, 60, ...] $\times$ $\sigma$, see the respective panels for values of $\sigma$.
The beam in the left panel is 1\farcs13 $\times$ 0\farcs93, and the beam in the right
panel is 0\farcs45 $\times$ 0\farcs33. 
}
\end{figure}

\begin{figure}[!ht]
\begin{center}
\includegraphics[scale=0.6]{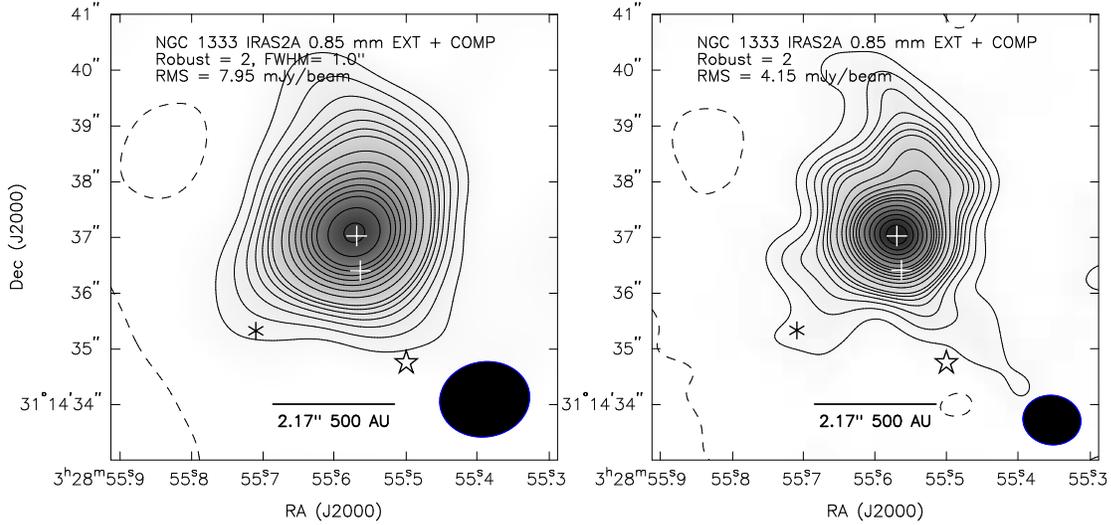}
\end{center}
\caption{NGC 1333 IRAS2A system as observed by the SMA at 850 \micron. A lower resolution image
is reconstructed from a combination of Compact and Extend configuration data 
with Robust = 2 weighting and 1.0\arcsec\ tapering is shown
in the left panel. A higher resolution image is shown in the right panel Robust = 2 weighting and
no tapering. The crosses mark the positions of VLA1 and VLA2 derived from the 9 mm
images in Figure 1.
Like Figure 2, these images also do not show evidence of
discrete sources at the locations of MM2 and MM3 marked by the asterisk and star symbols respectively.
MM2/MM3 have expected detection levels of $\sim$9.7$\sigma$/17.3$\sigma$ in the higher resolution image. 
The contour levels are in the left and right panels are [-3, 3, 5, 7, 9, 11, 13,\
15, 20, 25, 30, 35, 40, 45, 50, 60, ...] $\times$ $\sigma$, see the 
respective panels for values of $\sigma$. 
The beam in the left panel is 1\farcs04 $\times$ 0\farcs89, and the beam in the right
panel is 0\farcs87 $\times$ 0\farcs60.}
\end{figure}

\begin{figure}[!ht]
\begin{center}
\vspace{20mm}
\includegraphics[trim=0.5cm 12cm 0.5cm 4cm,scale=0.6]{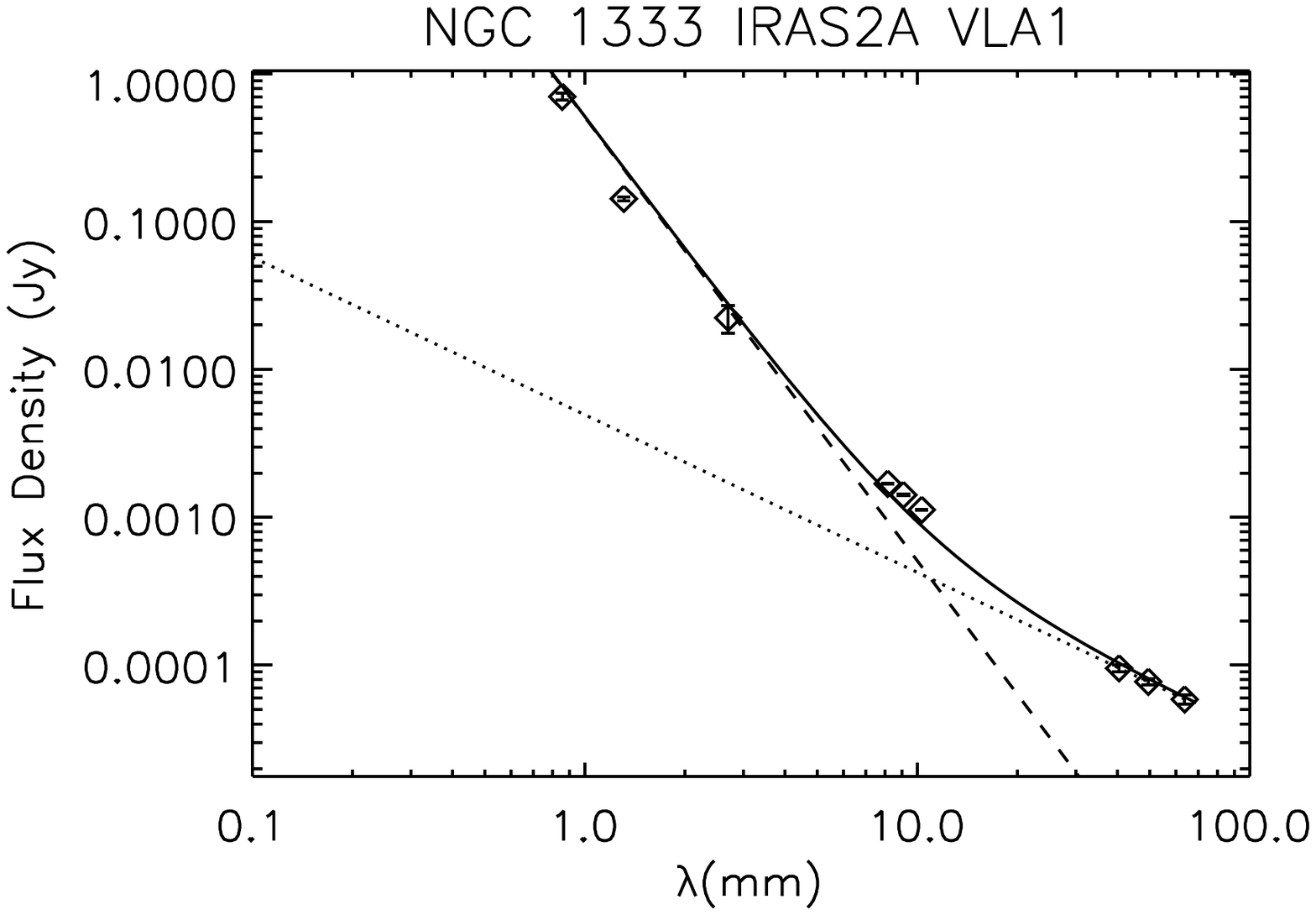}
\includegraphics[trim=0.5cm 12cm 0.5cm 4cm,scale=0.6]{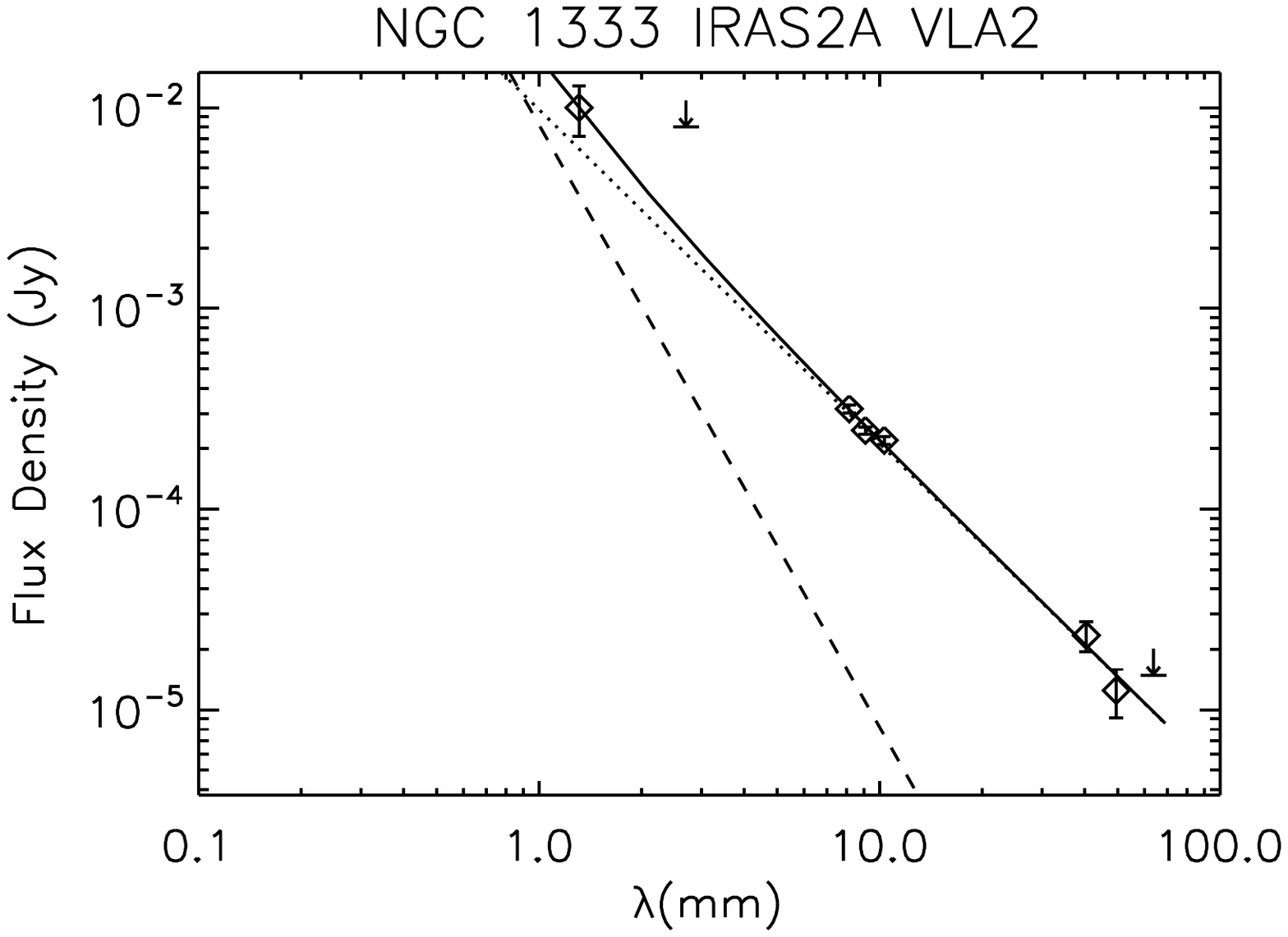}
\end{center}
\caption{Millimeter to centimeter-wave spectra of IRAS2A VLA1 and VLA2 using the integrated intensity
data from Table 1. The
dotted line is a fit to the free-free slope 
(using only the 4 cm and 6.4 cm data for VLA1),
 the dashed line is a fit to the thermal dust emission slope,
and the solid line is the sum of the thermal dust emission and free-free component. The 2.7 mm points for
VLA1 and VLA2 are the integrated flux density and 3$\sigma$ upper limits from \citet{looney2000};
the 2.7 mm flux of VLA1 is used in the fit. The 4 cm
and 4.9 cm points for VLA2 are the peak flux densities rather than the integrated flux due to low signal-to-noise.
For unresolved sources, like VLA2, the peak flux density is equivalent to the integrated flux.}
\label{sedplots}
\end{figure}

\begin{figure}[!ht]
\begin{center}
\includegraphics[scale=0.6]{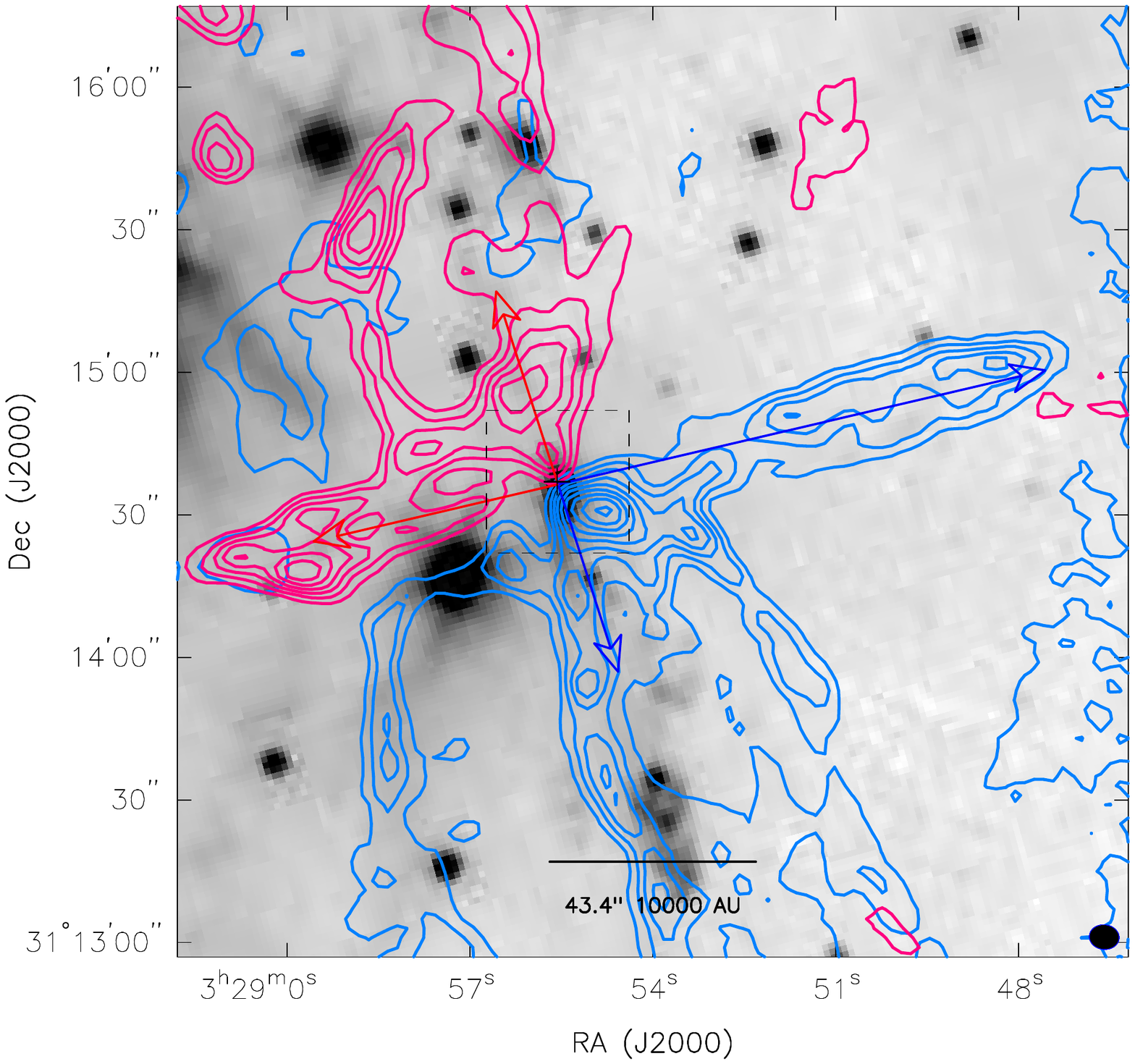}
\end{center}
\caption{Blue-shifted and red-shifted CO ($J=1\rightarrow0$) contours from \citet{plunkett2013}
overlaid on the \textit{Spitzer} 4.5 \micron\ image (grayscale) from \citet{jorgensen2006}. 
The cross marks the position of VLA1 derived from the 9 mm
image in Figure 1.
The CO emission traces the large-scale
outflow emission associated with the IRAS2A system, showing the quadrupolar outflow from the 
source. The contours are [-10, 10, 20, 30, ...] $\times$ $\sigma$ and $\sigma$ = 0.61 K and 0.76 K for the
blue and red integrated intensity maps, respectively. The beam is 6\farcs3 $\times$ 5\farcs2. The CO emission is integrated
over 0.5 \kms\ to 6.5 \kms\ (blue contours) and 9.7 \kms\ to 13.5 \kms\ (red contours). The dashed box
shows the region that viewed in CO ($J=2\rightarrow1$) and ($J=3\rightarrow2$) in Figure 6.}
\label{IRAS2A-outflow-large}
\end{figure}

\begin{figure}[!ht]
\begin{center}
\includegraphics[scale=0.6]{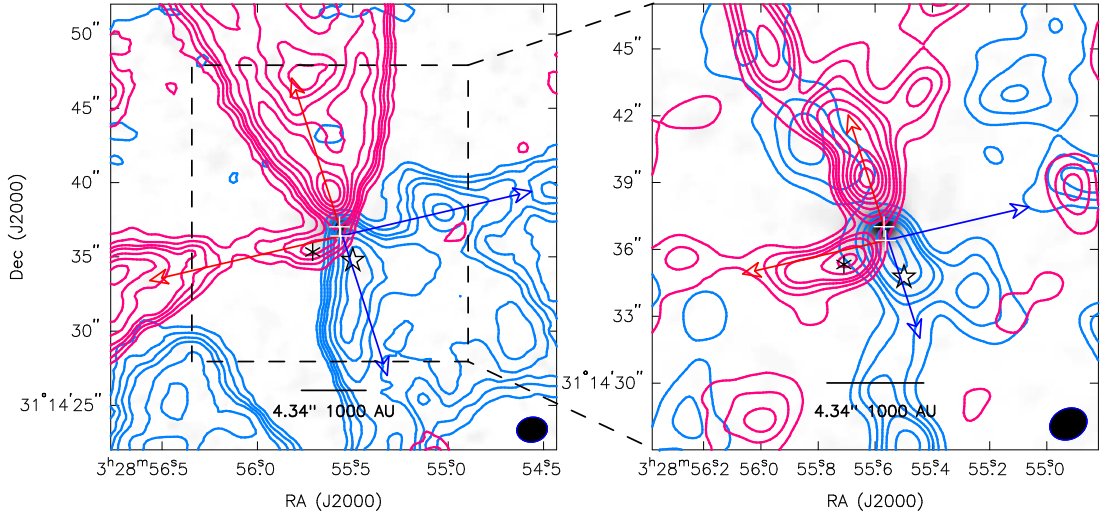}

\end{center}
\caption{Blue-shifted and red-shifted CO ($J=2\rightarrow1$) from a 
combination of CARMA C, D, and E-array observations with Robust = 0 weighting (left)
and SMA CO ($J=3\rightarrow2$) from a combination of Compact and Extended configurations with
Natural weighting (right). Both outflow maps are overlaid on the CARMA 1.3 mm continuum image (grayscale).
The crosses mark the positions of VLA1 and VLA2 derived from the 9 mm
images in Figure 1.
These higher excitation CO lines, observed at higher resolution, trace the outflowing material 
closer to the protostars and show that there are departures from outflow symmetry on small-scales
relative to the larger scale outflow axis. The asterisk and star symbols mark
the positions of MM2 and MM3, respectively, from \citet{codella2014}. Note that MM2 
lies exactly in a bright spot of CO emission traced by both transitions, and MM3 is in the
blue-shifted lobe of north-south outflow. The contours are [-5, 5, 10, 15, 20, 30, ...] $\times$ $\sigma$ 
for the red and blue-shifted emission in both the left and right panels. The RMS noise values are 
0.81 K, 0.74 K, 0.88 K, 0.77 K for the blue and red CO ($J=2\rightarrow1$) integrated intensity maps
in the left panel and the blue and red  CO ($J=3\rightarrow2$) integrated intensity maps in the right panel, respectively.
The beams are 2\farcs1 $\times$ 1\farcs7 in the left panel and 1\farcs7 $\times$ 1\farcs4 in the right panel.
The CO ($J=2\rightarrow1$) emission is integrated over 2.75 \kms\ to 5.75 \kms\ (blue contours) and 9.75 \kms\ to 12.5 \kms\ (red contours);
the CO ($J=3\rightarrow2$) emission is integrated over 1.0 \kms\ to 5.75 \kms\ (blue contours) and 9.7 \kms\ to 13.0 \kms\ (red contours).
}
\label{IRAS2A-outflow-small}
\end{figure}

\clearpage
\begin{deluxetable}{lllllllllll}
\tablewidth{0pt}
\rotate
\tabletypesize{\tiny}
\tablecaption{NGC 1333 IRAS2A Measurements}
\tablehead{
  \colhead{Wavelength} & \colhead{Array} & \colhead{Integrated Flux Density} & \colhead{Peak Flux Density}  & \colhead{Gaussian Size} & \colhead{Gaussian PA} & \colhead{Deconvolved Size} & \colhead{Deconvolved PA} & \colhead{Robust}  & \colhead{Beam}      & \colhead{Beam PA}\\
  \colhead{(mm)}        & \colhead{Config.}    & \colhead{(mJy)}           & \colhead{(mJy/beam)}      & \colhead{(\arcsec)}     & \colhead{(\degr)}     & \colhead{(\arcsec)}        & \colhead{(\degr)}        &                   & \colhead{(\arcsec)} & \colhead{(\degr)} \\
}
\startdata
VLA1\\
0.85 & EXT+COMP & 701 $\pm$ 38 & 361 $\pm$ 4.6    & 1.15 $\times$ 0.99   & 87.5  & 0.81  $\times$ 0.74  & 152   & -2  & 0.87 $\times$ 0.60 & 84.5\\
1.3  & BC  & 143.0 $\pm$ 4.5  & 96.6 $\pm$ 1.5      & 0.56 $\times$ 0.42   & 134.9 & 0.34  $\times$ 0.26  & 147   & 1  & 0.45 $\times$ 0.33 & -50.7\\ 
8.15 & AB  & 1.69 $\pm$ 0.014 & 1.43 $\pm$ 0.0084   & 0.13 $\times$ 0.11   & 112.2 & 0.054 $\times$ 0.036 & 110.6 & 2  & 0.11 $\times$ 0.10 & -66.7\\
8.15 & A   & 1.73 $\pm$ 0.019 & 1.23 $\pm$ 0.013    & 0.090 $\times$ 0.071 & 107.9 & 0.055 $\times$ 0.033 & 109.0 & 2  & 0.07 $\times$ 0.06 & -73.9\\
9.1  & AB  & 1.42 $\pm$ 0.012 & 1.23 $\pm$ 0.0056   & 0.14 $\times$ 0.12   & 111.9 & 0.056 $\times$ 0.038 & 113.9 & 2  & 0.12 $\times$ 0.11 & -69.6\\
9.1  & A   & 1.44 $\pm$ 0.012 & 1.09 $\pm$ 0.0086   & 0.096 $\times$ 0.079 & 111.29 & 0.054 $\times$ 0.032 & 111.6 & 2 & 0.08 $\times$ 0.07 & -69.3\\
10.3 & AB  & 1.13 $\pm$ 0.01  & 0.99 $\pm$ 0.0071   & 0.15 $\times$ 0.13   & 112.0 & 0.058 $\times$ 0.036 & 117.8 & 2  & 0.13 $\times$ 0.12 & -72.7\\
10.3 & A   & 1.14 $\pm$ 0.014 & 0.91 $\pm$ 0.012    & 0.10 $\times$ 0.085  & 117.0 & 0.054 $\times$ 0.030 & 117.9 & 2  & 0.09 $\times$ 0.08 & -64.7\\
40.5 & A & 0.096 $\pm$ 0.005 & 0.079 $\pm$ 0.0038   & 0.39 $\times$ 0.33   & 92.1  & 0.18  $\times$ 0.11  & 125   & 2  & 0.35 $\times$ 0.30 & 78.4\\
49.6 & A & 0.077 $\pm$ 0.004 & 0.066 $\pm$ 0.0034   & 0.47 $\times$ 0.38   & 76.9  & 0.23 $\times$ 0.11   & 79    & 2  & 0.41 $\times$ 0.36 & 75.2\\
63.8 & A & 0.0588 $\pm$ 0.004 & 0.054 $\pm$ 0.005   & 0.68 $\times$ 0.43   & 72.3  & ...                  & ...   & 2  & 0.55 $\times$ 0.47 & 78.5\\

VLA2\\
1.3\tablenotemark{a}  & BC & 10.0  $\pm$ 2.8   & 9.7   $\pm$ 1.5     & ... & ...   & ...                  & ... & 1   & 0.45 $\times$ 0.33 & -50.7\\ 
8.15 & AB & 0.316 $\pm$ 0.014 & 0.276 $\pm$ 0.0084  & 0.12 $\times$ 0.11   & 97.6  & 0.049 $\times$ 0.032 & 39  & 2   & 0.11 $\times$ 0.10 & -66.7\\
8.15 & A & 0.328 $\pm$ 0.017 & 0.248 $\pm$ 0.013  & 0.082 $\times$ 0.071   & 91.5  & 0.042 $\times$ 0.031 & 68  & 2   & 0.07 $\times$ 0.06 & -73.9\\
9.1  & AB & 0.247 $\pm$ 0.01  & 0.245 $\pm$ 0.0059  & 0.12 $\times$ 0.11   & 106.1 & ...                  & ... & 2   & 0.12 $\times$ 0.11 & -69.6\\
9.1  & A & 0.263 $\pm$ 0.01  & 0.229 $\pm$ 0.0086  & 0.084 $\times$ 0.076   & 83.8 & ...                  & ... & 2   & 0.08 $\times$ 0.07 & -69.3\\
10.3 & AB & 0.220 $\pm$ 0.01  & 0.213 $\pm$ 0.0071  & 0.14 $\times$ 0.12   & 102.5 & ...                  & ... & 2   & 0.13 $\times$ 0.12 & -72.7\\
10.3 & A & 0.221 $\pm$ 0.012  & 0.203 $\pm$ 0.012  & 0.091 $\times$ 0.082   & 62.4 & ...                  & ... & 2   & 0.09 $\times$ 0.08 & -64.7\\
40.5 & A  & 0.015 $\pm$ 0.004 & 0.023 $\pm$ 0.0038  & ...                  & ... & ...                  & ... & 2   & 0.35 $\times$ 0.30 & 78.4\\
49.6 & A  & 0.008 $\pm$ 0.003 & 0.0125 $\pm$ 0.0034 & ...                 & ... & ...                  & ... & 2   & 0.41 $\times$ 0.36 & 75.2\\
63.8 & A  & $<$0.0015         &  $<$0.0015          & ...                   & ...  & ...                  & ...   & 2  & 0.55 $\times$ 0.47 & 78.5\\
\enddata
\tablecomments{All flux measurements are from Gaussian fits to the images at each wavelength, except for the peak flux densities
which are directly taken from the images. We do not give uncertainties on the Gaussian fit parameters for the sake of brevity.
The 1$\sigma$ RMS noise for each image can be found in the uncertainty of the peak flux density. Note that the wavelengths
are listed with higher precision than in the text.}
\tablenotetext{a}{Due to the faintness of the source we fixed the source parameters to match the beam and we only allowed the flux to vary.}

\end{deluxetable}

\begin{deluxetable}{lllll}
\tablewidth{0pt}
\tabletypesize{\scriptsize}
\tablecaption{IRAS2A Spectral Slopes}
\tablehead{
  \colhead{} & \colhead{VLA1}   & \colhead{VLA1} & \colhead{VLA2} & \colhead{VLA2 ($\lambda$ $>$ 7 mm)}\\
  \colhead{} & \colhead{(fixed free-free)}   & \colhead{(no fixed parameters)} & \colhead{(fixed thermal slope)} & \colhead{(no fixed parameters)}\\}
\startdata
Free-free Slope     &  1.1$\pm$0.19  & 0.8$\pm$0.4   &  1.7$\pm$0.12  & 1.7$\pm$0.09\\
F$_0$ (free-free)   &  4.9$\pm$3.5    & 1.5$\pm$2.5    &  9.8$\pm$3.2    &  10.2$\pm$2.0\\
Thermal Dust Slope  &  2.8$\pm$0.08  & 2.6$\pm$0.14  &  3.0            &  ...\\
F$_0$ (dust)        & 379$\pm$59      & 316$\pm$60     &  8.2$\pm$1.0    &  ...\\

\enddata
\tablecomments{The spectral slopes are defined by the convention 
F$_{\lambda}$ = F$_{0}(\lambda/\lambda_0)^{-\alpha}$ where
$\alpha$ is the spectral slope and $\lambda_0$ = 1 mm. The VLA2 ($\lambda$ $>$ 7 mm) values
are from a single slope fit to the $\lambda$ $>$ 7 mm data, obtaining
a similar slope to the fit with a fixed thermal slope.}

\end{deluxetable}

\end{document}